\documentclass[lettersize,journal]{IEEEtran}
\usepackage{amsmath,amsfonts}
\usepackage[ruled,linesnumbered]{algorithm2e}
\usepackage{array}
\usepackage[caption=false,font=normalsize,labelfont=sf,textfont=sf]{subfig}
\usepackage{textcomp}
\usepackage{stfloats}
\usepackage{url}
\usepackage{verbatim}
\usepackage{graphicx}
\def\BibTeX{{\rm B\kern-.05em{\sc i\kern-.025em b}\kern-.08em
		T\kern-.1667em\lower.7ex\hbox{E}\kern-.125emX}}
\usepackage{balance}
\usepackage{diagbox}
\usepackage{cite}
\usepackage{amssymb}
\usepackage{color}

\begin{document}
	
	
\title{Spectral Comb Shaping \textcolor{black}{for Single Carrier Communication Signals} by Polar Codes}
\author{Yinuo Mei, and Daiming Qu
	\thanks{This work was supported in part by the National Science Foundation of China (No. 62331010). (Corresponding author: Daiming
		Qu.)
		
		The authors are with the School of Electronic Information and
		Communications, Huazhong University of Science and Technology,
		Wuhan 430074, China. (email: D202280923@hust.edu.cn; qudaiming@hust.edu.cn).}}

\maketitle

\begin{abstract}
An approach to selecting information indices for
polar codes is proposed to form signals with spectral comb shapes under
BPSK modulation, whereby the signal could be separated from periodic 
interference in spectrum. By confining information indices to an index set termed comb-shaping index set (CIS) proposed in this paper, a spectral comb shape signal is
formed, which has periodic nulls and notch bands in its spectrum. \textcolor{black}{Furthermore, we propose a novel construction for polar coding under the CIS constraint.}
Numerical results are given under periodic interference and AWGN noise,
indicating that a considerable signal-to-noise power ratio (SNR) gain is
accomplished in comparison with conventional polar codes.
\end{abstract}


\section{Introduction}

\IEEEPARstart{I}{n} a growing number of modern communication scenarios, from radio-frequency (RF) communication to sonar applications \cite{ref:Harmonic_RF,ref:Harmonic_sonar}, from ultra wideband (UWB) modulations to Internet-of-Things (IoT) systems \cite{ref:Harmonic_UWB,ref:Harmonic_IoT}, signals must contend with a particularly stubborn form of impairment: strong, periodic interference. The prevalence and high power of such interference pose a significant bottleneck to achieving reliable, high-speed data transmission in these critical environments. The conventional approach to combatting periodic interference is to employ a comb filter at the receiver\cite{ref:Analog Comb Filter,ref:Digital Comb Filter}. However, this simplicity comes at a steep and often unacceptable price: signal distortion. Through a comb filter, along with the periodic interference, the components of signals on interference frequencies are also removed, which causes signal distortion. In summary, simply filtering is a passive and destructive strategy.

This fundamental limitation of passive filtering motivates a proactive shaping of the signal's spectrum. Specifically, instead of simply filtering out interference at the receiver, if we proactively shape the transmitted signal's spectrum to introduce nulls precisely at the frequencies where periodic interference exists, the signal could be spectrally separable from the interference, thus not suffering distortion caused by comb filters. Here what is required is a concrete shaping scheme for signals to introduce periodic spectral nulls, namely, spectral comb shaping.

{\color{black}
In multi-carrier communications like OFDM \cite{ref:OFDM}, by nulling specific carrier frequencies, spectral comb shaping could be achieved. However, in single carrier communications, there is no such straightforward method. Compared to multi-carrier communications like OFDM, single carrier ones exhibit significantly lower PAPR, which enables high power amplifier efficiency and extended coverage \cite{ref:SinCrr1}; moreover, single carrier schemes offer lower implement complexity. The merits and prospects of single carrier communications render them an indispensable class of communication scenarios \cite{ref:SinCrr1,ref:SinCrr2}. Therefore, in this work, we confine our investigation to spectral comb shaping in single carrier communications.
} 

According to Fourier transform, there is a relationship between temporal
periodicity and spectral-null periodicity, which is going to be further
discussed and extended in this paper. \textcolor{black}{Spread-spectrum schemes, like CDMA \cite{ref:CDMA1,ref:CDMA2}, have the potential to generate periodic codes by properly designing the spreading code. However, by this means, the temporal periodicity has to be achieved within each single symbol period, namely the symbol rate should be no higher than the interference fundamental frequency, thereby strictly limited. To attain higher symbol rate, we resort to channel coding to achieve the required temporal periodicity}, where polar coding is a prominent candidate and thereby considered in this paper. \textcolor{black}{Unlike unstructured codes such as LDPC \cite{ref:LDPC}, whose generator matrix is random and irregular,} polar codes are highly structured \cite{ref:Polar}. The inherent recursive structure of polar coding, stemming from the Kronecker power of the generator matrix, provides a natural foundation for enforcing temporal periodicity in codewords, rendering it an ideal candidate for spectral comb shaping.

{\color{black}
Polar coding with specific additional constraints has been discussed in some contexts. In order to constrain the codeword probability distribution to a target low-cost distribution, a probabilistic shaping scheme for polar codes is proposed in \cite{ref:Polar cnst1} by configuring dynamic frozen bits. As a contrast, our research is aimed at constructing polar codes with the constraint of temporal periodicity, which could lead to a comb-like spectrum after modulation.
} 

Furthermore, from the perspective of error correction, polar codes \cite{ref:Polar}, as the first kind of deterministic capacity-achieving codes \cite{ref:Lim} that can be constructed, encoded, and decoded with acceptable complexity, have been applied to various communication fields and have remarkable performance \cite{ref:Polar app1,ref:Polar app2,ref:Polar app3}. Therefore, the employment of polar codes for signal spectral comb-shaping excels in both periodic interference resistance and error correction, which would have a promising communication performance on noisy channels with periodic interference.

In this paper, we propose an approach to selecting appropriate information indices to generate signals with periodic spectral nulls, which could be separated from periodic interference in the spectrum. \textcolor{black}{The selection would confine information indices to an index set termed comb-shaping index set (CIS) proposed in this paper, whose size is half the code length, therefore, source bits indexed outside the CIS, namely half of the source bits, are fixed as zeros. As a consequence, the prior assumption of conventional polar codes that all source bits are i.i.d. uniformly distributed over \(\{0,1\}\) ceases to hold. Therefore, the sub-channel capacity under CIS constraint is not equal to the symmetric sub-channel capacity, i.e., the sub-channel capacity while source bits are i.i.d. uniformly distributed over \(\{0,1\}\). Namely there is a discrepancy between CIS-constrained sub-channel capacity and symmetric sub-channel capacity. To solve this discrepancy, we further propose a novel construction for the proposed polar code, whereby the correct CIS-constrained sub-channel capacity is obtained and the best indices are allocated to information bits, thus leading to a performance enhancement in error correction.}

The rest of this paper is organized as follows. In Section II, basic
concepts over periodic interference, polar codes and BPSK modulation are
presented. Section III investigates what kind of signals have periodic spectral nulls, and corresponding codewords under BPSK modulation,
then discusses the selection of polar generator matrix rows to generate
such codewords. Furthermore, the solution of capacity discrepancy 
is discussed in Section IV. The performance of frame error rate is compared between proposed scheme and
conventional one in Section V. Finally, Section VI serves as a summary
of above contents.

\section{Preliminaries}

This paper focuses on polar coding, BPSK modulation and endeavors to resist the periodic interference, therefore the preliminaries of them are presented below.

\subsection{Periodic Interference and Comb Filters}

Periodic interference is interference whose spectrum takes nonzero values only at a
fundamental frequency and its harmonic frequencies. In other words, the
support set of its spectrum are periodic frequency points. The
expression of the spectrum is
\begin{align}
	I(f) = \sum_{k = - \infty}^{\infty}{a_{k}\delta(f-kf_I)}
\end{align}
where \(f_I\) is the interference fundamental frequency, and \(\delta( \cdot )\) represents the impulse function. Namely the interference frequency set, denoted by \(\Phi\), is \(\{kf_I|k\in\mathbb{Z}\}\).

However, actual periodic interference may not have such ideal form, it
could have a bandwidth around each harmonic frequency, namely the
spectrum of it could be
\begin{align} \label{eq_actItf}
	I(f) = \sum_{k = - \infty}^{\infty}{I_{k}(f)}
\end{align}
where \(I_k\) \textcolor{black}{is the \(k\)-th interference tone,} which has a bandwidth of \(B_I\) centered at \(kf_I\), \textcolor{black}{and \(B_I\) represents the bandwidth of interference tones, which could be termed interference tone-bandwidth}.

To remove periodic interference, a filter whose frequency response exactly takes zero values at interference frequencies 
is required. Namely
\begin{align}
	H(kf_I)=0,k\in \mathbb{Z}
\end{align}

The regular notch bands on its frequency response make it seem like a comb, so such kind of filters are termed comb filters \cite{ref:Digital Comb Filter}.

\subsection{Polar Codes}

Polar codes are a class of linear block codes. Encoding of polar codes
with code length \(N\) and information bit number \(K\), namely code
rate \(R = \frac{K}{N}\), includes assigning information bits
\(c_{0}^{K - 1}\) to an \emph{N}-length bit sequence \(u_{0}^{N - 1}\),
and mapping source bits \(u_{0}^{N - 1}\) to final codeword
\(x_{0}^{N - 1}\) with a linear operator, which can be represented as a
matrix \(G_{N}\) termed generator matrix. In expressions, codeword
\(x_{0}^{N - 1} = u_{0}^{N - 1}G_{N}\), where \(u_{0}^{N - 1}\) is a
combination of information bits \(c_{0}^{K - 1}\) and frozen bits zeros,
namely
\(u_{\mathcal{A}} = c_{0}^{K - 1},u_{\mathcal{A}^{\complement}} = \lbrack 0,\ldots,0\rbrack\).
Here \(\mathcal{A}\) is the information index set, namely the index set allocated to
information bits, contrarily bits at indices
\(\mathcal{A}^{\complement} = \left\{ 0,1,\ldots,N - 1 \right\}\mathcal{\backslash A}\) in \(u_{0}^{N - 1}\) are zeros. Namely,
\(x_0^{N-1}=u_{\mathcal{A}} G_N(\mathcal{A})\), where \(G_N\left( \mathcal{A} \right)\)
refers to the submatrix of \(G_N\) composed of its rows indexed in
\(\mathcal{A}\). Generator matrix of polar codes with code
length \(N = 2^{m},m \in \mathbb{N}^{*}\) is
\(G_{N} = B_{N}F^{\otimes m}=F^{\otimes m}B_{N}\), where \(B_{N}\) is a bit-reversal
permutation matrix, and \(F = \begin{bmatrix}
1 & 0 \\
1 & 1
\end{bmatrix}\) is termed polar kernel, besides, operator \(\otimes n\)
means \emph{n}-th Kronecker power \cite{ref:Polar}.

Then the codeword \(x_0^{N-1}\) is transmitted through a memoryless channel \(W^N\), which represents the \(N\) uses of binary-input channel \(W\), and the output of \(W^N\) is \(y_0^{N-1}\). While decoding, polar code utilizes received signal and former source
bits to estimate each source bit, specifically, it constructs \(N\)
channels for every source bit, termed sub-channels, the \emph{i}-th of
which is
\(U_{i} \mapsto Y_{0}^{N - 1}U_{0}^{i - 1},i \in \left\{ 0,1,\ldots,N - 1 \right\}\).
So the transition probability of the \emph{i}-th sub-channel is
\(W_{N}^{(i)}\left( y_{0}^{N - 1},u_{0}^{i - 1} \middle| u_{i} \right)\).
Due to the recursive property of generator matrix \(G_{N}\), on
memoryless channels, it is not difficult to derive recursive expression
between transition probability on adjacent layers \cite{ref:Polar}

{\small
\begin{align*}
	&W_{2N}^{(2i)}\left( y_{0}^{2N - 1},u_{0}^{2i - 1} \middle| u_{2i} \right)= \\
	&\sum_{u_{2i + 1}\in\mathbb{B}}^{}{W_{N}^{(i)}\left( y_{0}^{N - 1},v_{0,0}^{i - 1} \middle| u_{2i} + u_{2i + 1} \right)W_{N}^{(i)}\left( y_{N}^{2N - 1},v_{1,0}^{i - 1} \middle| u_{2i + 1} \right)} \\
	&W_{2N}^{(2i + 1)}\left( y_{0}^{2N - 1},u_{0}^{2i} \middle| u_{2i + 1} \right) \\
	&= W_{N}^{(i)}\left( y_{0}^{N - 1},v_{0,0}^{i - 1} \middle| u_{2i} + u_{2i + 1} \right)W_{N}^{(i)}\left( y_{N}^{2N - 1},v_{1,0}^{i - 1} \middle| u_{2i + 1} \right)
\end{align*}
}where \(v_{0,i} \mathord{=} u_{2i} + u_{2i + 1},v_{1,i} \mathord{=} u_{2i + 1},\forall i \in \left\{ 0,1,\ldots,N - 1 \right\}\), and \(\mathbb{B}:=\{0,1\}\).

The fundamental decoding algorithm of polar code is successive
cancellation (SC) \cite{ref:Polar}. SC estimates each source bit \(u_{i}\) by
received signal \(y_{0}^{N - 1}\) and former estimated source
bits \({\widehat{u}}_{0}^{i - 1}\), and estimation at index \(i\) viz.
\({\widehat{u}}_{i}\) is exploited along with former estimations
\({\widehat{u}}_{0}^{i - 1}\) in subsequent decoding, as following
expression indicates:
\begin{align*}
	{\widehat{u}}_{i} = \left\{
	\begin{aligned}
		& \arg{\max_{u_{i} \in \mathbb{B}}{W_{N}^{(i)}\left( y_{0}^{N - 1},{\widehat{u}}_{0}^{i - 1} \middle| u_{i} \right)}} & ,i \in \mathcal{A} \\
		& 0 & ,i \in \mathcal{A}^{\complement}
	\end{aligned} \right.\ 
\end{align*}

Moreover, I. Tal and A. Vardy proposed a list decoding algorithm, which
saves multiple decoding paths during decoding, each of which records an
estimated bit sequence. Such algorithm is successive cancellation list (SCL) \cite{ref:SCL}. Considerable
performance and acceptable time and space complexity render it a practical
decoding algorithm for polar codes.

\subsection{BPSK}

Binary phase shift keying (BPSK) is a classic kind of binary modulation,
in which bits are mapped to constellation symbols \(q(x_{n})\), and the
latter are sent to a shaping filter to generate the modulated signal
\(s(t)\) as
\begin{align}
	s(t) = \sum_{n = 0}^{N - 1}{q\left( x_{n} \right)p\left( t - nT_{b} \right)}
\end{align}
where \(x_{0}^{N - 1}\in\mathbb{B}^N\) is a binary sequence with length \(N\), \(q:\mathbb{B}\mapsto\mathbb{C}\) is
the BPSK mapping, \(T_{b}\) is the symbol period, and \(p\) is the
shaping pulse with finite duration, e.g., rectangular pulse, squared root raised cosine (SRRC) pulse, etc. \cite{ref:BPSK}. The symbol rate is \(R_s:=\frac{1}{T_b}\). In other words, denote a BPSK
modulator by \(\mathcal{M}_{\mathrm{BPSK}}\), then
\begin{align}
	\mathcal{M}_{\mathrm{BPSK}}\left( x_{0}^{N - 1} \right) = \sum_{n = 0}^{N - 1}{q\left( x_{n} \right)p\left( t - nT_{b} \right)}
\end{align}

In the receiver, after matched filtering and down sampling, we
get the received symbols \(y_{0}^{N - 1}\), where
\(y_{n} = \frac{1}{\lVert p \rVert ^2} \left( r(t)*p^*(-t) \right)(nT_{b})\). Therefore
\(y_{0}^{N - 1} = q\left( x_{0}^{N - 1} \right) + z_{0}^{N - 1}\),
\(z_{0}^{N - 1}\) is AWGN noise. Here
\(q\left( x_{0}^{N - 1} \right) := \left\lbrack q\left( x_{n} \right) \right\rbrack_{n \in \left\{ 0,\ldots,N - 1 \right\}}\).
In other words, BPSK with an AWGN channel is equivalent to a mapper
\(q\) with an AWGN channel.

\section{Periodic Spectral Nulls Introduced by Polar Codes}

While filtering out periodic interference, in order to avoid signal distortion caused by a comb filter, we should guarantee that the signal and periodic interference are separable in spectrum, namely the signal spectral notches cover all interference tones. Since interference frequencies are fixed, it is necessary to introduce periodic
nulls in the signal spectrum to cover these frequencies. To begin
with, the relationship between temporal periodicity and spectral nulls
periodicity is to be discussed. Then the structure of polar codes
will be exploited to achieve such temporal periodicity and introduce periodic
spectral nulls.

\subsection{Regularly-Repetitive Signals and Locally-Periodic Sequences}

We define a particular kind of signals termed regularly-repetitive signals below:

\emph{Definition 1:} Regularly-repetitive Signals. A signal \(s(t)\) is
a regularly-repetitive signal if there exists a signal \(s_{0}\) and
\(M\in\mathbb{N} \backslash \mathbb{B}, T\in\mathbb{R}\) that
\begin{align}
	s(t) = \sum_{m = 0}^{M - 1}{s_{0}(t - mT)}
\end{align}

For a regularly-repetitive signal, \(T\) is termed repetition interval
(RI), and \(M\) is termed repetition number (RN).

Then for regularly-repetitive signals we have Theorem 1.

\emph{Theorem 1:} A regularly-repetitive signal with RI \(T\) and RN
\(M\) has spectral nulls
\(k\frac{1}{M}f_{s},k\in\mathbb{Z} \backslash M\mathbb{Z}\), where
\(f_{s} = \frac{1}{T}\).

\emph{Proof of Theorem 1:} For a regularly-repetitive signal \(s(t)\),
according to Definition 1,
\(s(t) = \sum_{m = 0}^{M - 1}{s_{0}(t - mT)}\), therefore its spectrum
\(S(f)\) is
\begin{align*}
	&S(f) = \mathcal{F}\left\lbrack s(t) \right\rbrack\mathcal{= F}\left\lbrack \sum_{m = 0}^{M - 1}{s_{0}(t - mT)} \right\rbrack \nonumber \\
	&= \sum_{m = 0}^{M - 1}{\mathcal{F}\left\lbrack s_{0}(t - mT) \right\rbrack} = \sum_{m = 0}^{M - 1}{\mathcal{F}\left\lbrack s_{0}(t) \right\rbrack e^{- j2\pi fmT}} \nonumber \\
	&= \sum_{m = 0}^{M - 1}{S_{0}(f)e^{- j2\pi fmT}} = S_{0}(f)\sum_{m = 0}^{M - 1}e^{- j2\pi fmT} \nonumber \\
	&= S_{0}(f)G_{M,T}(f)
\end{align*}

Function \(G_{M,T}(f) := \sum_{m = 0}^{M - 1}e^{- j2\pi fnT}\), a direct calculation yields
\[G_{M,T}(f) = \begin{cases}
	\frac{\sin{\pi MTf}}{\sin{\pi Tf}}e^{- j(M - 1)T\pi f} & ,f \neq kf_{s} \\
	M & ,f = kf_{s}
\end{cases} \]

Trivially, the nulls of \(G_{M,T}(f)\) are \(k\frac{1}{M}f_{s},k \in \mathbb{Z} \backslash M\mathbb{Z}\). Therefore, \(S(f)\) has nulls
\(k\frac{1}{M}f_{s},k \in \mathbb{Z} \backslash M\mathbb{Z}\). \(\hfill \blacksquare\)

The exploitation of regularly-repetitive signals to introduce periodic spectral
nulls constitutes the main idea of this paper. Furthermore, in
this paper we focus on BPSK-modulated signals, whose regular-repetitiveness could be
attributed to a corresponding property of the binary sequences. Hereby
we propose a property of sequences termed \emph{local-periodicity}, and
give the definition of locally-periodic sequences in Definition 2.

\emph{Definition 2:} Locally-Periodic Sequences. A sequence
\(\xi_{0}^{N - 1}\) is a locally-periodic sequence if it is a
succession of multiple subsequences with identical length and period,
i.e.,
\(\exists M \in \mathbb{N}\backslash \mathbb{B}, L \in \mathbb{N}^{*},K \in \mathbb{N}^{*}\)
that \(N = KML\) and
\(\forall k \in \left\{ 0,\ldots,K - 1 \right\},l \in \left\{ 0,\ldots,L - 1 \right\},m \in \left\{ 0,\ldots,M - 1 \right\}\)
there is
\begin{align}
	\xi_{kML + l} = \xi_{kML + l + mL}
\end{align}

For a locally-period sequence, \(L\) is termed local period (LP), and
\(M\) is termed local period number (LPN).

For example, sequence 0110 0110 0110 1011 1011 1011 1100 1100 1100 0010
0010 0010 0001 0001 0001 is a locally-periodic sequence with LP 4 and
LPN 3, because it is a succession of five subsequences: 0110 0110 0110,
1011 1011 1011, 1100 1100 1100, 0010 0010 0010 and 0001 0001 0001, each
of which is a periodic sequence with period 4 and has 3 periods.

It is not difficult to prove that the BPSK signal of a locally-periodic
binary sequence is regularly-repetitive, which is recorded in Lemma 1.

\emph{Lemma 1:} For a locally-periodic binary sequence with LP \(L\) and LPN
\(M\), \textcolor{black}{and for arbitrary shaping pulse,} its BPSK-modulated signal is regularly-repetitive with RI
\(LT_{b}\) and RN \(M\), where \(T_{b}\) is the symbol period.

The proof of Lemma 1 is straightforward, thereby omitted. Intuitively, by mapping a symbol to a pulse, the local-periodicity of binary sequences translates into regularly-repetitiveness of signals.

{\color{black}
\subsection{Comb Shaping by Spread-Spectrum and Its Limitation}
	
As discussed above, local-periodicity of binary sequences translates into signal regular-repetitiveness and periodic spectral nulls. Then an immediate idea for spectral comb-shaping is an application of spread-spectrum schemes. For spread-spectrum-based shaping schemes, such as CDMA \cite{ref:CDMA1,ref:CDMA2}, since the symbols are random, the local-periodicity is supposed to be achieved within each single symbol period, namely by a design of the spreading code. This constraint leads to a strict limitation of the symbol rate \(R_s\).

Specifically, under spreading factor \(P\), by choosing a periodic spreading code with two periods, according to Definition 2, the chip sequence is locally-periodic with LP \(\frac{1}{2} P\) and LPN \(2\), therefore according to Lemma 1, the signal is regularly-repetitive with RI \(\frac{1}{2} T_s\) and RN \(2\), then according to Theorem 1 it has spectral nulls \(k R_{s},k \in \mathbb{Z} \backslash 2\mathbb{Z}=1+2\mathbb{Z}\), i.e., periodic spectral nulls \((1+2a)R_s,a \in \mathbb{Z}\). Therefore, to ensure the spectral nulls cover all interference frequencies, the symbol rate ought to be no higher than half of the interference fundamental frequency, i.e., \(R_s \leq \frac{1}{2} f_I\), which severely restricts the communication rate. For example, while the interference fundamental frequency is 50Hz, the symbol rate should be no higher than 25Hz, thereby severely limited. It is a crucial limitation of spread-spectrum-based schemes.

In this paper, to attain higher and more flexible symbol rate, we resort to channel coding to introduce periodic spectral nulls.
}

\subsection{Rows of Polar Generator Matrix with Local-Periodicity}

\begin{figure*}[!t]
	\centering
	\includegraphics[width=7in]{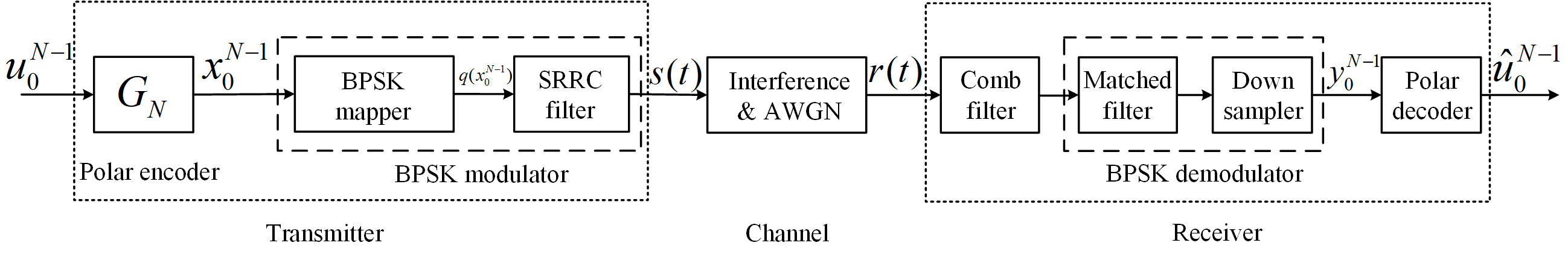}
	\caption{The system model.}
	\label{fig_sys}
\end{figure*}

The system model of this research is presented in Fig.~\ref{fig_sys}, where the
information bits are polar encoded and BPSK-modulated with a SRRC filter to generate a
signal, which would suffer from periodic interference and AWGN on the
channel. In the receiver, BPSK demodulator and polar decoder are
applied to resist the AWGN noise, but before them a comb filter is used
to remove the periodic interference, which asks for a spectral
separability between the signal and periodic interference to avoid
distortion, as discussed in Section I. 

From Section II-B we know in polar encoding \(x_0^{N-1}=\sum_{k\in\mathcal{A}}{c_{k}G_N(k)}\), where \(G_N(k)\) denotes the \(k\)-th row of polar generator matrix \(G_N\). Therefore, if specific rows of polar generator matrix are
locally-periodic, then by confining the information indices \(\mathcal{A}\)
to these rows' indices, the codewords could be locally-periodic, and through
BPSK modulation the signals could be regularly-repetitive. In this subsection, we will further explore the local-periodicity of
rows of polar generator matrix.

The generator matrix of polar codes is \(G_N=B_N F^{\otimes m}\), and \(F^{\otimes m}\) is a Kronecker
power of Arikan's kernel \(F = \begin{bmatrix}
1 & 0 \\
1 & 1
\end{bmatrix}\). The Kronecker power
has an iterative structure that would cause a repetition of some bits,
so it is plausible to suppose that some rows of \(F^{\otimes m}\) are
locally-periodic, namely some rows of \(G_N\) are locally-periodic, because \(B_N\) is merely a permutation. An elaborated discussion of the local-periodicity
is given below.

According to Kronecker power, \(F^{\otimes m} = F \otimes F_{m - 1}\), and
according to Kronecker product there is
\(F^{\otimes m}(i,j)
= F\left( \left\lfloor \frac{i}{2^{m - 1}} \right\rfloor,\left\lfloor \frac{j}{2^{m - 1}} \right\rfloor \right)F_{m - 1}\left( R_{2^{m - 1}}(i),R_{2^{m - 1}}(j) \right)\)
, and recalling \(F = \left[\begin{smallmatrix}
	1 & 0 \\
	1 & 1
\end{smallmatrix}\right]\), i.e., \(F(i,j) = !(! i\& j)\), there is
\(F^{\otimes m}(i,j) = \mathord{!}\left( !\left\lfloor \frac{i}{2^{m - 1}} \right\rfloor\&\left\lfloor \frac{j}{2^{m - 1}} \right\rfloor \right)\& F_{m - 1}\left( R_{2^{m - 1}}(i),R_{2^{m - 1}}(j) \right)\), where \(!\) represents the NOT operation, \(\&\) represents the bitwise AND operation, and \(\lfloor \cdot \rfloor\) represents the floor function, \(R_n(k)\) represents the remainder of \(k\) divided by \(n\). 

For convenience, let \(i_d\) denote the \(d\)-th bit in the binary representation of \(i\), where
\(d \in \left\{ 0,\ldots m - 1 \right\}\), then there is
{\small\[F^{\otimes m}(i,j) = !\left( ! i_{m - 1}\& j_{m - 1} \right)\& F_{m - 1}\left( i_{0}^{m - 2},j_{0}^{m - 2} \right)\]}so iteratively, we have
{\small
\begin{align*}
	&F^{\otimes m}(i,j) \\
	&= !\left( ! i_{m - 1}\& j_{m - 1} \right)\&!\left( ! i_{m - 2}\& j_{m - 2} \right) \& F_{m - 2}\left( i_{0}^{m - 3},j_{0}^{m - 3} \right) \\
	&= \ldots \\
	&= !\left( ! i_{m - 1}\& j_{m - 1} \right)\&!\left( ! i_{m - 2}\& j_{m - 2} \right)\& \ldots \&!\left( ! i_{0}\& j_{0} \right) \\
	&= \&_{d = 0}^{m - 1}\left\lbrack !\left( ! i_{d}\& j_{d} \right) \right\rbrack = ! |_{d = 0}^{m - 1}\left( ! i_{d}\& j_{d} \right)
\end{align*}
}where \(|\) represents the bitwise OR operation. Note that \(G_N=F^{\otimes m}B_N\), i.e., \(G_N(i,j)=F^{\otimes m}(i,\sum_{d=0}^{m-1}{j_{m-d-1} \cdot 2^d})\), there is
\begin{align} \label{eq_eleGN}
	G_N(i,j) = ! |_{d = 0}^{m - 1}\left( ! i_d \& j_{m-d-1} \right)
\end{align}

According to formula (\ref{eq_eleGN}) and the truth table of the bitwise AND operation \&, if any \(i_{d}\)
is 1, then the value of \(j_{m-d-1}\) does not affect the value of
\(G_N(i,j)\). From such discovery we have Lemma 2.

\emph{Lemma 2:} For any index
\(i \in \left\{ 0,\ldots,2^{m} - 1 \right\}\), if the \(r\)-th digit
of \(i\) is 1, i.e., \(i_r = 1\), then the \(i\)-th row of polar
generator matrix \(G_N\), i.e., \(G_N(i)\) is locally-periodic with
LP \(2^{m-r-1}\) and LPN 2.

\emph{Proof of Lemma 2:} If \(\ i_r = 1\), then for any two
\(j,j^{\prime} \in \left\{ 0,\ldots,2^{m} - 1 \right\}\) that
\(j_{d} \oplus j_{d}^{\prime} = !\left( d - (m-r-1) \right)\), according to
formula (\ref{eq_eleGN}) there is
\begin{align*}
	&G_N(i,j) = !\left\{ \left\lbrack |_{d \in \left\{ 0,\ldots, m - 1 \right\}\backslash r}\left( ! i_{d}\& j_{m-d-1} \right) \right\rbrack | \left( ! i_r \& j_{m-r-1} \right) \right\} \\
	&= !\left\{ \left\lbrack |_{d \in \left\{ 0,\ldots, m - 1 \right\}\backslash r}\left( ! i_{d}\& j_{m-d-1} \right) \right\rbrack | 0 \right\} \\
	&= !\left\{ \left\lbrack |_{d \in \left\{ 0,\ldots, m - 1 \right\}\backslash r}\left( ! i_{d}\& j_{m-d-1}^{\prime} \right) \right\rbrack | \left( ! i_r \& j_{m-r-1}^{\prime} \right) \right\} \\
	&= G_N\left( i,j^{\prime} \right)
\end{align*}
where \(\oplus\) represents the XOR operation, and \(!\) represents the
NOT operation.

For any \(j \in \left\{ 0,\ldots,2^{m} - 1 \right\}\) that
\(j_{m-r-1} = 0\), trivially
\(j_{d} \oplus \left( j + 2^{m-r-1} \right)_{d} = !\left( d - (m-r-1) \right)\),
therefore, \(G_N(i,j) = G_N\left( i,j + 2^{m-r-1} \right)\). Then
\(\forall k \in \left\{ 0,\ldots,2^r - 1 \right\},l \in \left\{ 0,\ldots,2^{m-r-1} - 1 \right\}\),
since
\(\left( 2k \cdot 2^{m-r-1} + l \right)_{m-r-1} = (2k)_{0} = 0\), with
above conclusion there is
\(G_N\left( i,2k \cdot 2^{m-r-1} + l \right) = G_N\left( i,2k \cdot 2^{m-r-1} + l + 2^{m-r-1} \right)\),
so \(\forall m^{\prime} \in \mathbb{B}\) there is
\begin{align*}
	G_N\left( i,k \cdot 2 \cdot 2^{m-r-1} + l + m^{\prime} \cdot 2^{m-r-1} \right) \\
	= G_N\left( i,k \cdot 2 \cdot 2^{m-r-1} + l \right)
\end{align*}

Then let \(M = 2\), \(L = 2^{m-r-1}\) and \(K = 2^r\),
above equation is
\(G_N\left( i,kML + l + m^{\prime}L \right) = G_N(i,kML + l)\), according
to Definition 2, \(G_N(i)\) is a locally-periodic sequence with LP
\(2^{m-r-1}\) and LPN 2. \(\hfill \blacksquare\)

Therefore, if we define an index set with
parameter \(r\) by
\(\Lambda_{r} = \left\{ i \middle| i \in \left\{ 0,\ldots,2^{m} - 1 \right\},i_r = 1 \right\}\),
then an index set is established, polar generator matrix rows indexed within which are locally-periodic with the same LP and LPN. Later we will further prove that by confining information indices to a
\(\Lambda_{r}\), the BPSK-modulated signal of the polar codeword could
have periodic spectral nulls, therefore, we refer to \(\Lambda_{r}\) as the \(r\)-th
\textit{comb-shaping index set} (\textit{CIS}), and a row of polar generator matrix
indexed in the \(r\)-th CIS, i.e., \(G_N(i),i \in \Lambda_{r}\) as an
\(r\)-th comb-shaping row (CR), as Definition 3:

\emph{Definition 3:} Comb-Shaping Index Sets (CIS) and Comb-Shaping Rows
(CR). For a polar code at length \(N=2^{m},m \in \mathbb{N}^{*}\), the
\(r\)-th comb-shaping index set (CIS) is defined as
\begin{align}
	\Lambda_{r} := \left\{ i \middle| i \in \left\{ 0,\ldots,2^{m} - 1 \right\},i_r = 1 \right\}
\end{align}
in other words
\begin{align}
	\Lambda_{r} := \bigcup_{k = 0}^{2^{m-r-1} - 1}\left\{ (2k + 1) \cdot 2^r,\ldots,(2k + 2) \cdot 2^r - 1 \right\}
\end{align}
and any \(G_N(i),i \in \Lambda_{r}\) is termed an \(r\)-th comb-shaping
row (CR) of the polar generator matrix \(G_N\). Here
\(r \in \left\{ 0,\ldots,m - 1 \right\}\) is termed the CIS order.

Moreover, a polar code employing a CIS, namely selecting information indices within a CIS, is termed a \textit{comb-shaping polar code}.

\begin{figure}[!t]
	\centering
	\includegraphics[width=3.5in]{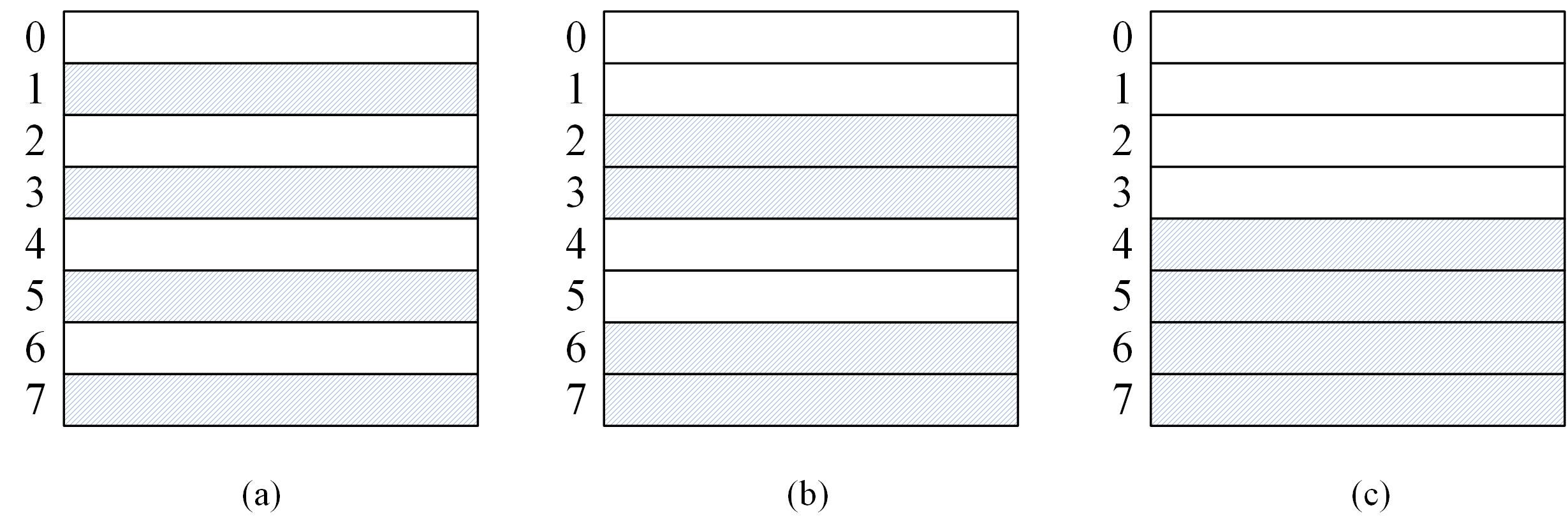}
	\caption{Rows of \(G_8\) indexed in different \(\Lambda_{r}\). (a)\(\ r = 0\); (b) \(r = 1\); (c) \(r = 2\).}
	\label{fig_CIS}
\end{figure}

For example, while CIS order \(r = 0\), CIS \(\Lambda_{r}\) is
\(\Lambda_{0} = \left\{ 1,3,5,\ldots,N-3,N - 1 \right\}\);
while \(r = 1\),
\(\Lambda_{1} = \left\{ 2,3,6,7,\ldots,N-2,N-1 \right\}\);
and while \(r = 2\),
\(\Lambda_{2} = \left\{4,5,6,7,12,13,14,15,\ldots,N-4,N-3,N-2,N-1\right\}\), etc. Fig.~\ref{fig_CIS} presents rows of \(G_8\) indexed in different \(\Lambda_{r},r\in\{0,1,2\}\), which are marked by shadows.

With the definition of term CIS, it is immediate to obtain following theorem:

\emph{Theorem 2:} For a polar generator matrix \(G_N\), where \(N=2^m\), if \(i\) is an index belonging to the \(r\)-th CIS, i.e., \(i \in \Lambda_{r},r\in\{0,\ldots,m-1\}\), then the \(i\)-th row if \(G_N\), i.e., \(G_N(i)\) is locally-periodic with LP \(2^{m - r - 1}\) and LPN 2.

\emph{Proof of Theorem 2:} For any \(i \in \Lambda_{r}\),
\(r \in \left\{ 0,\ldots,m - 1 \right\}\), according to Definition 3, there is
\(i_r = 1\), then according to Lemma 2, \(G_N(i)\) is
locally-periodic with LP \(2^{m - r - 1}\) and LPN 2. \(\hfill \blacksquare\)

\begin{figure}[!t]
	\centering
	\includegraphics[width=3.5in]{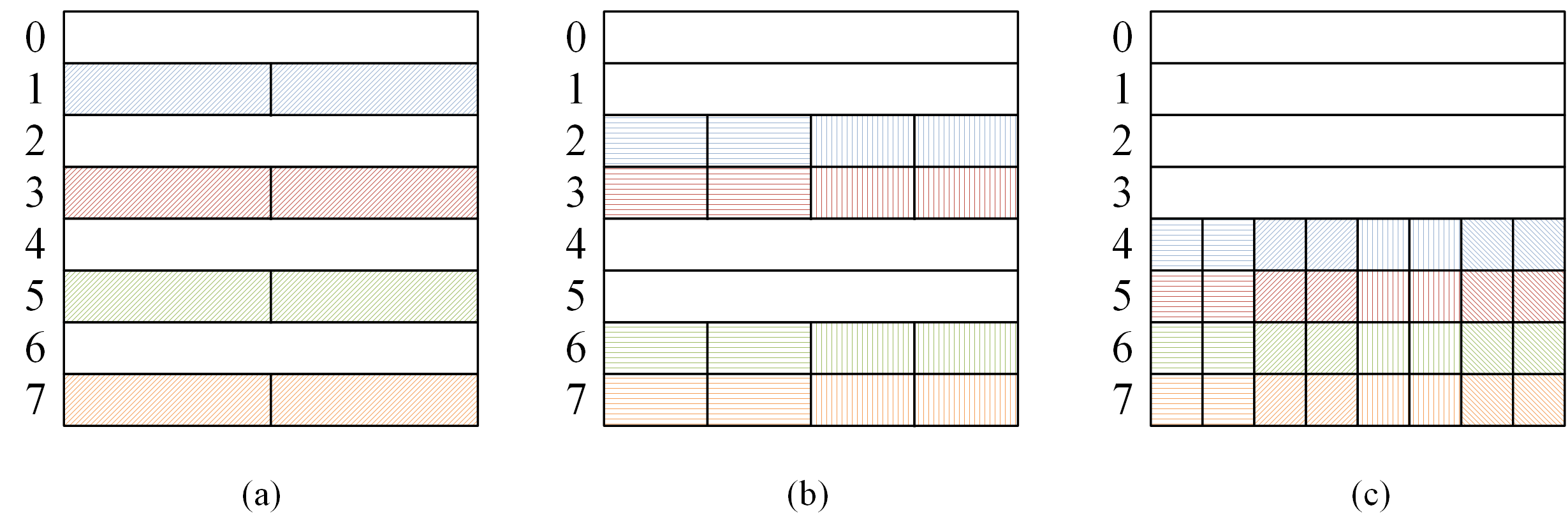}
	\caption{Local-periodicity of \(G_8\) rows indexed in different \(\Lambda_{r}\). (a)\(\ r = 0\); (b) \(r = 1\); (c) \(r = 2\).}
	\label{fig_LP}
\end{figure}

As an instance, Fig.~\ref{fig_LP} depicts the local-periodicity of \(G_8\) rows indexed in different \(\Lambda_{r},r\in\{0,1,2\}\). In each row,
shadows of the same pattern indicate identical subsequence.

If the information index
set \(\mathcal{A}\) is a subset of \(\Lambda_{r}\),
\(r \in \left\{ 0,\ldots,m - 1 \right\}\), then the codeword
\(x_{0}^{N - 1} = u_{\mathcal{A}}G_N\left( \mathcal{A} \right)\), as a
linear combination of rows in \(\Lambda_{r}\), is locally-periodic with
LP \(2^{m - r - 1}\) and LPN 2. Through BPSK modulation the signal is
regularly-repetitive, thus having periodic spectral nulls. The following
subsection would specifically present the periodic spectral nulls introduced
by provided comb-shaping rows.

\subsection{Periodic Spectral Nulls Introduced with a CIS}

Now, we can finally connect this structural property of the code back to the spectral properties of the transmitted signal. By ensuring all information bits are loaded onto rows from the same CIS, the resulting codeword inherits this periodicity, which, as we will prove in Theorem 3, translates directly into predictable nulls in the frequency domain.

\emph{Theorem 3:} For a polar code at length \(N = 2^{m},m\in\mathbb{N}^*\) and a BPSK modulation \textcolor{black}{with arbitrary shaping pulse}, if the
information index set \(\mathcal{A}\) is a subset of \(\Lambda_{r}\),
i.e., \(\mathcal{A \subseteq}\Lambda_{r}\), the BPSK-modulated signal of
the codeword has spectral-null set
\begin{align}
	\Theta_{r} = \left\{ (1 + 2a) \cdot 2^{r}f_{w} \middle| a\in\mathbb{Z} \right\}
\end{align}

In other words
\begin{align}
	\mathcal{\forall A \subseteq}\Lambda_{r},u_{\mathcal{A}} \in \mathbb{B}^{K},a \in \mathbb{Z}, S\left( (1 + 2a) \cdot 2^{r}f_{w} \right) = 0
\end{align}
where
\(S(f) := \mathcal{F}\left\lbrack \mathcal{M}_{\mathrm{BPSK}}\left( u_{\mathcal{A}} G_N(\mathcal{A}) \right) \right\rbrack\),
\(f_{w} := \frac{1}{NT_{b}}\), \(T_{b}\) is the symbol period.

\emph{Proof of Theorem 3:} \(\forall \mathcal{A} \subseteq \Lambda_{r}\),
each index
\(\mathcal{A}_{k} \in \Lambda_{r},k \in \left\{ 0,\ldots,\left| \mathcal{A} \right| - 1 \right\}\),
so according to Theorem 2, \(G_N\left( \mathcal{A}_{k} \right)\) is
locally-periodic with LP \(2^{m - r - 1}\) and LPN 2. Furthermore, the
codeword
\(x_{0}^{N - 1} = u_{\mathcal{A}} G_N(\mathcal{A}) = \sum_{k = 0}^{K - 1}{c_{k}G_N\left( \mathcal{A}_{k} \right)}\),
so \(x_{0}^{N - 1}\) is locally-periodic with LP
\(2^{m - r - 1}\) and LPN 2. Then for
\(s(t) = \mathcal{M}_{\mathrm{BPSK}}(x_{0}^{N - 1})\), i.e., the BPSK-modulated
signal of \(x_{0}^{N - 1}\), according to Lemma 1, it is
regularly-repetitive with RI \(2^{m - r - 1} \cdot T_{b}\) and RN 2,
where \(T_{b}\) is the symbol period. Finally with Theorem 1 its
spectrum \(S(f)\) has nulls
\(k\frac{1}{2}\frac{1}{2^{m - r - 1} \cdot T_{b}} = k \cdot 2^{r}\frac{1}{NT_{b}} = k \cdot 2^{r}f_{w},k \in \mathbb{Z}\backslash2\mathbb{Z =}1 + 2\mathbb{Z}\),
namely \(S(f)\) has nulls
\((1 + 2a) \cdot 2^{r}f_{w},a\in\mathbb{Z}\). \(\hfill \blacksquare\)

From the spectral-null set \(\Theta_{r}\), we know the period of spectral nulls is \(2^{r+1}f_w\), where \(f_w\) is termed codeword frequency. Therefore, a physical significance of
CIS order \(r\) is revealed: \(r\) is the base-2 logarithm of the
semiperiod of spectral nulls normalized by the codeword frequency,
in other words, \(2^{r}\) is the semiperiod of spectral nulls
normalized by the codeword frequency.

{\color{black}
To introduce periodic spectral nulls, the information index set should be a subset of a CIS. According to Definition 3 we know the size of a CIS is half the code length, i.e., \(\frac{1}{2}N\). As a result, the code rate is upper bounded by \(\frac{1}{2}\).

Notably, the conclusion of spectral comb shaping by polar codes, i.e., Theorem 3 is crucially based on Theorem 2, which reveals the local-periodicity of specific polar codewords. Therefore, the essential property that enables polar codes to achieve spectral comb shaping is the structured coding. In contrast, unstructured codes like LDPC could not achieve the local-periodicity, thus not capable of spectral comb shaping.
} 

{\color{black}
\subsection{Parameter Determination for Signal-Interference Separability}

The purpose of signal-interference separability requires that the signal spectral notches cover all interference tones, which calls for a precise determination of parameters. According to Theorem 3, the baseband signal has spectral nulls \((1 + 2a) \cdot 2^{r}f_{w}, a\in\mathbb{Z}\); while the interference frequencies, according to Section II-A, are \(k f_I,k\in\mathbb{Z}\). To ensure that spectral notches of the bandpass signal cover all interference tones, it is necessary and sufficient that \(f_I\) is divisible by \(2^{r+1}f_w\), i.e., \(\frac{f_I}{2^{r + 1}f_{w}} \in \mathbb{N}^*\).

Trivially, iff \(\frac{f_I}{2f_w} \in \mathbb{N}^*\), there exists at least one \(r\in\{0,\ldots,m-1\}\) satisfying \(\frac{f_I}{2^{r + 1}f_{w}} \in \mathbb{N}^*\). Recall that \(f_w=\frac{R_s}{N}\), then \(\frac{f_I}{2f_w} \in \mathbb{N}^*\) is recast as
\begin{align} \label{eq_paraCons}
	\frac{N}{2R_s/f_I} \in \mathbb{N^*}
\end{align}

Formula (\ref{eq_paraCons}) specifies the constraint on the relationship among the symbol rate \(R_s\), the interference fundamental frequency \(f_I\) and the polar code length \(N\). In other words, provided that \(R_s\), \(f_I\) and \(N\) satisfy formula (\ref{eq_paraCons}), there exists at least one CIS that could be employed to generate signals spectrally separable from interference, namely the comb-shaping polar code could achieve signal-interference separability, which establishes a flexibility of the symbol rate and code length.

For instance, while \(f_I=50\mathrm{Hz}\) and the code length \(N\) is fixed at \(1024\), it is required that \(\frac{1024}{2R_s/f_I}\in\mathbb{N^*}\), i.e., \(\frac{2R_s}{f_I}=\frac{1024}{h},h \in \mathbb{N^*}\). Then \(R_s=\frac{1}{h} \cdot 25.6\mathrm{kHz}, h \in \mathbb{N^*}\), namely the symbol rate \(R_s\) could be 25.6kHz, 12.8kHz, 8.53kHz, 6.4kHz, 5.12kHz, 4.27kHz, 3.66kHz, 3.2kHz, \(\dots\) ; similarly, if fixing the symbol rate at \(3.2\mathrm{kHz}\), the code length is \(N=128h,h \in \mathbb{N^*}\), which could be 128, 256, 512, 1024, 2048, \(\dots\) .

Furthermore, it is highly desirable that spectral nulls exactly coincide with interference frequencies, thereby avoiding unnecessary spectral nulls, namely \(f_I=2^{r+1}f_w\), where \(r\) is the CIS order, i.e.
\begin{align} \label{eq_paraRecm}
	\frac{N}{2R_s/f_I} = 2^r,r\in\{0,\ldots,\log_2{N}-1\}
\end{align}

Therefore, it is recommended to set the symbol rate \(R_s\) as a power-of-two multiple of \(f_I\), and select a code length \(N\) larger than \(\frac{2R_s}{f_I}\). Then taking the CIS order by \(r=\log_2{\frac{N}{2R_s/f_I}}\), the baseband signals possess spectral nulls \((1+2a)\cdot \frac{1}{2}f_I,a\in\mathbb{Z}\). 

Finally, for periodic interference as presented by (\ref{eq_actItf}), the carrier frequency \(f_c\) should be chosen by \(f_c=(1+2b)\cdot \frac{1}{2}f_I\), where \(b\) could be an arbitrary natural number, then the spectral nulls of the bandpass signals are \(a\cdot f_I,a\in\mathbb{Z}\), which exactly coincide with interference frequencies.
} 

\begin{figure*}[!t]
	\centering
	\subfloat[]{\includegraphics[width=3.5in]{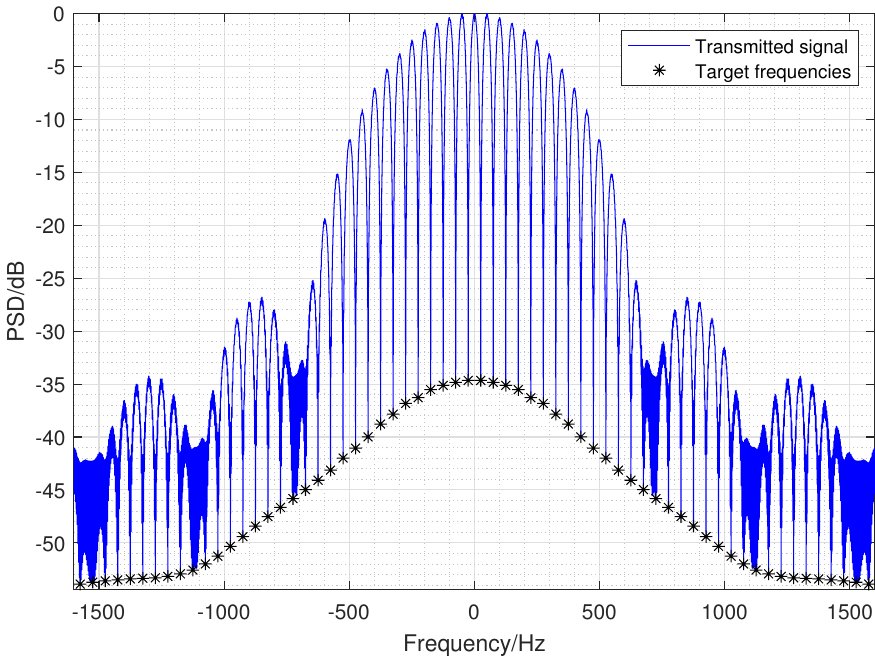}%
		\label{fig_first_case}}
	\hfil
	\subfloat[]{\includegraphics[width=3.5in]{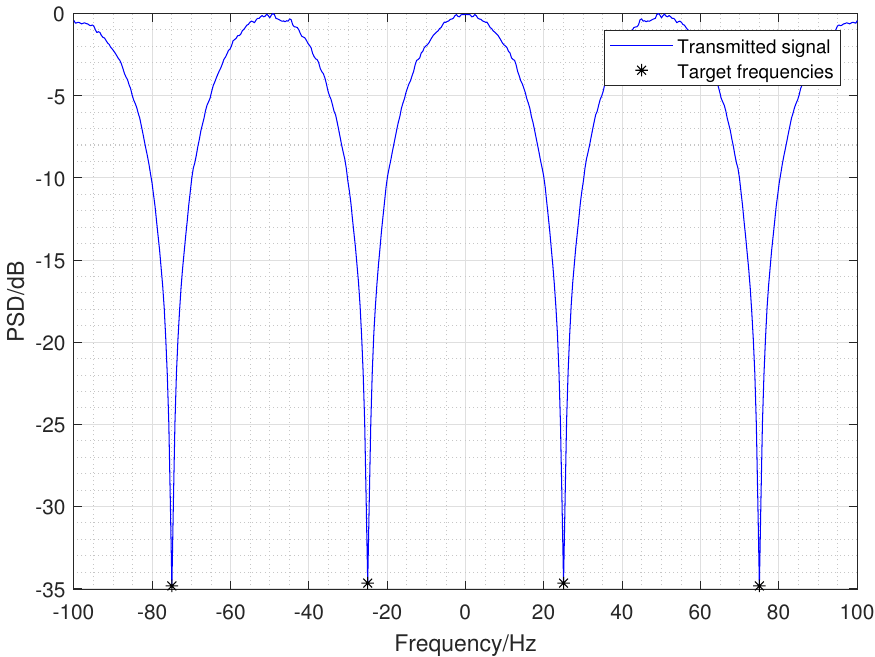}%
		\label{fig_second_case}}
	\caption{Power spectral density of the given example while
		\(r = 3\). (a) -1.6kHz to 1.6kHz; (b) -100Hz to 100Hz.}
	\label{fig_PSD}
\end{figure*}

For example, the interference fundamental frequency is
\(f_{I} = 50\mathrm{Hz}\), the symbol rate is
\(R_{s} = 800\mathrm{Hz}\), and the polar code length is \(N = 256\). Herein \(\frac{N}{2R_s/f_I}=8\in\mathbb{N^*}\), so formula (\ref{eq_paraCons}) is satisfied. Further \(\log_2{\frac{R_s}{f_I}}=4\in\{0,\ldots,7\}\), so taking \(r=\log_2{\frac{N}{2R_s/f_I}}=\log_2{8}=3\), formula (\ref{eq_paraRecm}) holds. Taking \(r=3\), the spectral nulls are \(\left\{ (1 + 2a) \cdot 2^{r}f_{w} \middle| a\in\mathbb{Z} \right\} = \left\{ (1 + 2a) \cdot \mathrm{25Hz} \middle| a\in\mathbb{Z} \right\} =\) \{..., -125Hz, -75Hz, -25Hz, 25Hz, 75Hz, 125Hz, ...\}. By ensuring the carrier frequency \(f_c=(1+2b)\cdot25\mathrm{Hz},b\in\mathbb{N}\), for example, 3.225kHz, the spectral nulls of the bandpass signal could cover all interference frequencies. In other words, the baseband frequencies corresponding to interference frequencies, which could be named target frequencies, are \{..., -125Hz, -75Hz, -25Hz, 25Hz, 75Hz, 125Hz, ...\}, which are covered by signal spectral nulls. While \(r=3\), taking sampling rate \(f_{s} = 6.4\mathrm{kHz}\), and an SRRC filter with roll-off factor 0.25, period number 2 and sampling factor 8, the power spectral density (PSD) of the signal is illustrated
in Fig.~\ref{fig_PSD}, where spectral nulls cover all target frequencies.

{\color{black}
In conclusion, by taking proper symbol rate and code length, and selecting information indices from a CIS with specific order, the purpose of signal-interference separability could be achieved.

Whilst the desired spectral shaping is successfully achieved by confining information indices to a CIS, it remains a question how to select information indices from a CIS. In the following section, we will discuss the capacity discrepancy caused by comb-shaping polar code, and propose a novel construction and decoding to address the problem.
} 

{\color{black}
\section{CIS-constrained Construction and Decoding of Polar Codes}

In conventional polar coding, it is assumed that all source bits \(U_0^{N-1}\) are i.i.d. uniformly distributed over \(\mathbb{B}\), in other words, the sub-channel prior distribution is \textit{symmetric} \cite{ref:Polar}. However, in comb-shaping polar coding, the information indices are confined to a CIS, whose size is half the code length, therefore, source bits outside the CIS, namely half of the source bits, are fixed as zeros. As a consequence, the prior assumption of conventional polar codes that all source bits are i.i.d. uniformly distributed over \(\mathbb{B}\) ceases to hold. This change will affect the sub-channel capacity and transition probability, therefore, a construction and decoding under CIS constraint is supposed to be designed for comb-shaping polar codes.

\subsection{CIS-constrained Construction}

In conventional polar coding, information index set selection is based on the symmetric capacity of sub-channels, i.e., sub-channel capacity under symmetric sub-channel prior distribution, namely the mutual information \(I(Y_0^{N-1}U_0^{i-1};U_i)\) while \(U_0^{N-1}\) are i.i.d. uniformly distributed over \(\mathbb{B}\) \cite{ref:Polar}. However, in comb-shaping polar coding, the information indices are confined to a CIS, whose size is half the code length, namely half of indices are frozen ones, which violates the prior assumption that each source
bit \(U_{i}\) is uniformly distributed over \(\mathbb{B}\), thereby
resulting in a \emph{capacity discrepancy}, namely the CIS-constrained sub-channel capacity is not
equal to the symmetric one, i.e., \(\forall r \in \left\{ 0,\ldots,m - 2 \right\},\exists i \in \Lambda_{r}\) that
\[I_{r,N}^{(i)} \neq I_{\mathrm{\mathrm{sym}},N}^{(i)}\]
where \(I_{r,N}^{(i)}\) represents the CIS-constrained capacity of the \(i\)-th sub-channel, i.e., the \(i\)-th sub-channel capacity
while the code length is \(N\) and source bits indexed outside CIS
\(\Lambda_{r}\) are zeros, indexed within \(\Lambda_{r}\) are i.i.d. uniformly
distributed over \(\mathbb{B}\); \(I_{\mathrm{sym},N}^{(i)}\) represents the symmetric capacity of the \(i\)-th sub-channel, i.e., the
\(i\)-th sub-channel capacity while the code length is \(N\) and
source bits are i.i.d. uniformly distributed
over \(\mathbb{B}\). Namely,
\(I_{r,N}^{(i)} := I\left( Y_{0}^{N - 1}U_{0}^{i - 1};U_{i} \right),U_{\Lambda_{r}^{\complement}} = \lbrack 0,\ldots,0\rbrack,U_{\Lambda_{r}}\sim\mathcal{B}^{N/2}\left( \frac{1}{2} \right),Y_{0}^{N - 1} = q\left( U_{0}^{N - 1}G_{N} \right) + Z_{0}^{N - 1}\);
\(I_{\mathrm{sym},N}^{(i)} := I\left( Y_{0}^{N - 1}U_{0}^{i - 1};U_{i} \right),U_{0}^{N - 1}\sim\mathcal{B}^{N}\left( \frac{1}{2} \right),Y_{0}^{N - 1} = q\left( U_{0}^{N - 1}G_{N} \right) + Z_{0}^{N - 1}\).
Here \(\mathcal{B}^{N}\left( \frac{1}{2} \right)\) denotes \(N\)-element
i.i.d. uniform distribution over \(\mathbb{B}\), and
\(Z_{0}^{N - 1}\) is AWGN.

As a consequence, under the CIS constraint, the symmetric sub-channel capacity is no longer suitable as a criterion for information index set selection.

Despite the capacity discrepancy at identical index, in Theorem 4 we have established a
relation between CIS-constrained capacities of \(\Lambda_{r}\) and symmetric
capacities at different indices:

\emph{Theorem 4:} In any CIS \(\Lambda_r,r\in\{0,\ldots,m-1\}\), at any index \(i \in \Lambda_r\), the CIS-constrained capacity of the \(i\)-th sub-channel, i.e., \(I_{r,N}^{(i)}\) equals the symmetric capacity of the \(g_{N,r}^{-1}(i)\)-th sub-channel:
\begin{align}
	I_{r,N}^{(i)} = I_{\mathrm{sym},N}^{(g_{N,r}^{-1}(i))}
\end{align}
where mapping \(g_{N,r}\) is defined by
\begin{align}
	g_{N,r}(i) := \left( 2\left\lfloor R_{N/2}(i)/2^{r} \right\rfloor + \left\lfloor i/(N/2) \right\rfloor \right)2^{r} + R_{2^{r}}(i)
\end{align}

The proof of Theorem 4 is in Appendix A. Here \(g_{N,r}\) is a mapping which maps indices within
\(\{ N/2,\ldots,N - 1\}\) into \(\Lambda_{r}\) and preserves their
order, i.e.,
\(g_{N,r}\left( \left\{ N/2,\ldots,N - 1 \right\} \right) = \Lambda_{r}\)
and
\(\forall i,j \in \Lambda_{r},i < j \rightarrow g_{N,r}(i) < g_{N,r}(j)\).
For example, mapping \(g_{16,1}\) is illustrated in Fig.~\ref{fig_mapping}. Its inverse mapping \(g_{N,r}^{-1}\) is
\(g_{N,r}^{- 1}(i) = \left\lfloor i/2^{r + 1} \right\rfloor 2^{r} + R_{2}\left( \left\lfloor i/2^{r} \right\rfloor \right)N/2 + R_{2^{r}}(i)\).
Trivially
\(g_{N,r}^{- 1}\left( \Lambda_{r} \right) = \left\{ N/2,\ldots,N - 1 \right\}\).
In other words, \(I_{r,N}^{\left( g_{N,r}(i) \right)} = I_{\mathrm{sym},N}^{(i)}\),
\(\forall i \in \left\{ N/2,\ldots N - 1 \right\}\).

\begin{figure}[!t]
	\centering
	\includegraphics[width=3.25in]{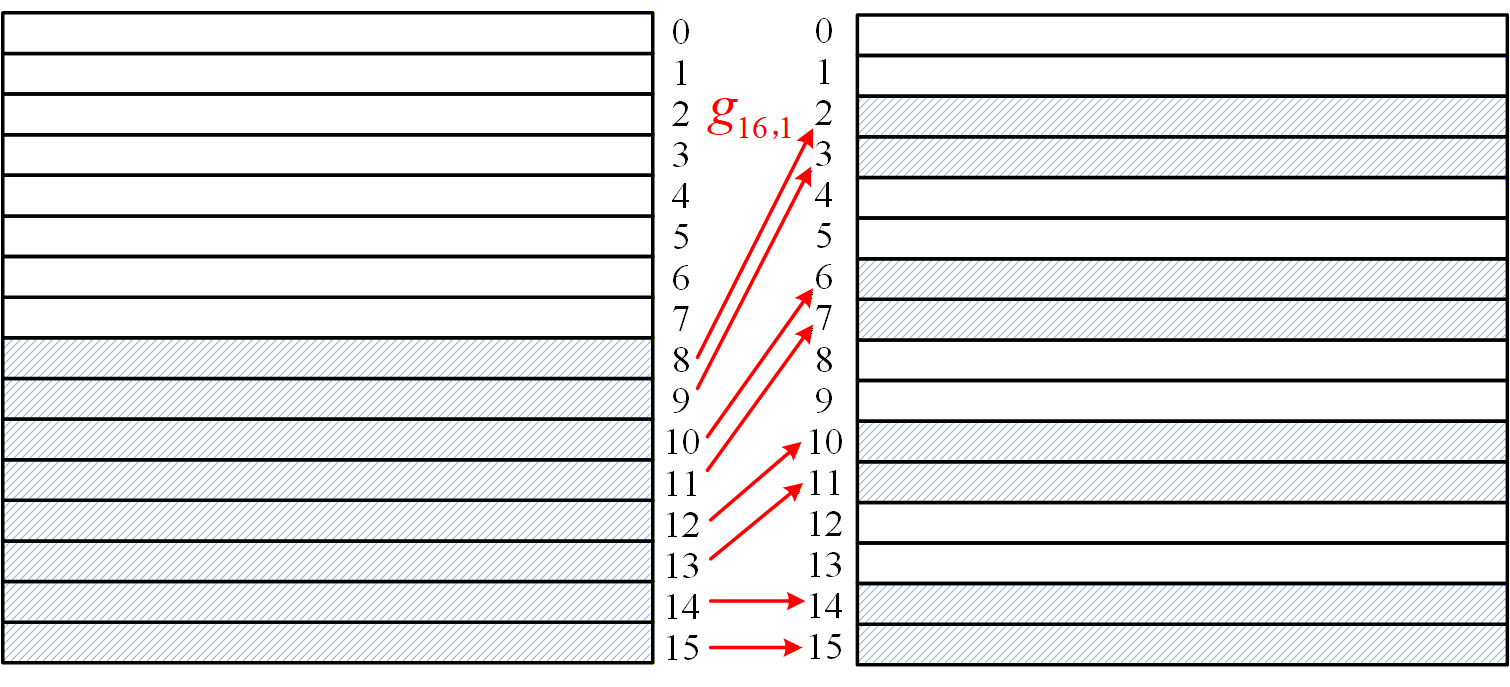}
	\caption{\textcolor{black}{Mapping \(g_{16,1}\).}}
	\label{fig_mapping}
\end{figure}

According to Theorem 4, by selecting indices with highest sub-channel
symmetric capacities within \(\left\{ N/2,\ldots,N - 1 \right\}\)
and mapping them into \(\Lambda_{r}\) by \(g_{N,r}\), the indices with
highest sub-channel CIS-constrained capacities within \(\Lambda_{r}\) are
determined, i.e.
\begin{align} \label{eq_sel}
	\mathcal{A} = g_{N,r}\left( \psi_{\mathrm{sym},K}\left( \left\{ N/2,\ldots,N - 1 \right\} \right) \right)
\end{align}
where function \(\psi_{\mathrm{sym},K}\) outputs \(K\) indices with highest sub-channel
symmetric capacities within the input index set. \(K\) is the number of information bits.

\begin{table}[]
	\begin{center}
		\caption{\textcolor{black}{Minimum CIS-constrained Sub-channel Capacity (MCSC) of the Information Index Set Selected under Different ISSC.}}
		\label{tab1}
		\begin{tabular}{|c|c|c|c|c|}
			\hline
			\diagbox{ISSC}{MCSC}{Code rate} & 1/4 & 5/16 & 3/8 \\
			\hline
			CIS-constrained-criterion & 0.9997 & 0.9901 & 0.8314  \\
			\hline
			Symmetric-criterion & 0.9984 & 0.9588 & 0.7627  \\
			\hline
		\end{tabular}
	\end{center}
\end{table}

We refer to the information index set selection criterion (ISSC) implied in formula (\ref{eq_sel}) as CIS-constrained-criterion, and the ISSC according to the symmetric capacity as symmetric-criterion.

With Theorem 4, CIS-constrained capacities of sub-channels indexed within each CIS could be obtained. For example, while the interference fundamental frequency is \(f_{I} = 50\mathrm{Hz}\), the symbol rate is
\(R_{s} = 800\mathrm{Hz}\), and the polar code length is \(N = 256\), as discussed in Section III-E, the preferred CIS order \(r\) is 3, so the CIS is \(\Lambda_3\). Table~\ref{tab1} shows minimum CIS-constrained sub-channel capacity (MCSC) of the information index set selected under different ISSC and different code rates. The signal-to-noise power ratio (SNR) of the AWGN noise, defined as the
power ratio of the signal to the noise within the signal band, is set as
-2dB. Table~\ref{tab1} illustrates that the MCSC of the information index set selected under CIS-constrained-criterion is significantly higher than the MCSC of the information index set selected under symmetric-criterion, which implies a superior performance of polar coding under CIS-constrained-criterion.

\subsection{CIS-constrained Decoding}

The violation of symmetric sub-channel prior distribution not only affects the sub-channel capacity, but also causes a \textit{transition probability discrepancy}, i.e., \(\forall r \in \left\{ 0,\ldots,m - 2 \right\},\exists i \in \Lambda_{r}\) that
\begin{align*}
	W_{r,N}^{(i)}\left( y_{0}^{N - 1},u_{\Lambda_{r}^{i - 1}} \middle| u_{i} \right)
	\neq W_{\mathrm{\mathrm{sym}},N}^{(i)}\left( y_{0}^{N - 1},u_0^{i - 1} \middle| u_{i} \right)
\end{align*}
where \(\Lambda_{r}^{i - 1}\) denotes
\(\Lambda_{r} \cap \left\{ 0,\ldots,i - 1 \right\}\), and
\(W_{r,N}^{(i)}\) represents the CIS-constrained transition probability of the \(i\)-th sub-channel, i.e., the \(i\)-th sub-channel transition probability while the code length is \(N\) and source bits indexed
outside CIS \(\Lambda_{r}\) are frozen, indexed within \(\Lambda_{r}\)
are i.i.d. uniformly distributed over \(\mathbb{B}\);
\(W_{\mathrm{sym},N}^{(i)}\) represents the CIS-constrained transition probability of the \(i\)-th sub-channel, i.e., the \(i\)-th sub-channel transition
probability while the code length is \(N\) and source bits are i.i.d.
uniformly distributed over \(\mathbb{B}\). Namely,
\(W_{r,N}^{(i)}\left( y_{0}^{N - 1},u_{\Lambda_{r}^{i - 1}} \middle| u_{i} \right) := p_{Y_{0}^{N - 1}U_{\Lambda_{r}^{i - 1}}|U_{i}}\left( y_{0}^{N - 1},u_{\Lambda_{r}^{i - 1}} \middle| u_{i} \right),U_{\Lambda_{r}^{\complement}} \mathord{=} \lbrack 0,\ldots,0\rbrack,U_{\Lambda_{r}} \sim \mathcal{B}^{N/2}\left( \frac{1}{2} \right),Y_{0}^{N - 1} = q\left( U_{0}^{N - 1}G_{N} \right) \mathord{+} Z_{0}^{N - 1}\);
\(W_{\mathrm{sym},N}^{(i)}\left( y_{0}^{N - 1},u_{0}^{i - 1} \middle| u_{i} \right) \mathord{:=} p_{Y_{0}^{N - 1}U_{0}^{i - 1}|U_{i}}\left( y_{0}^{N - 1},u_{0}^{i - 1} \middle| u_{i} \right), \\ U_{0}^{N - 1} \sim \mathcal{B}^{N}\left( \frac{1}{2} \right)\),\(Y_{0}^{N - 1} = q\left( U_{0}^{N - 1}G_{N} \right) + Z_{0}^{N - 1}\).
Here \(\mathcal{B}^{N}\left( \frac{1}{2} \right)\) denotes \(N\)-element
i.i.d. uniform distribution over \(\mathbb{B}\), and \(Z_{0}^{N - 1}\)
is AWGN.

The discrepancy in transition probability invalidates SCL decoding, which is based on symmetric prior distribution.

Nevertheless, through an appropriate transformation, the symmetric transition probability could still be used, which again enables SCL decoding. In Theorem 5 we establish the relation between the CIS-constrained
transition probability of \(\Lambda_{r}\) and symmetric one:

\emph{Theorem 5:} In any CIS \(\Lambda_r,r\in\{0,\ldots,m-1\}\), at any index \(i \in \Lambda_r\), the CIS-constrained transition probability \(W_{r,N}^{(i)}\) has following relation with the symmetric transition probability of the \(g_{N,r}^{-1}(i)\)-th sub-channel:
\begin{align}
	&W_{r,N}^{(i)}\left( y_{0}^{N - 1},u_{\Lambda_{r}^{i - 1}} \middle| u_{i} \right) \nonumber \\
	&= 2^{N/2}W_{\mathrm{\mathrm{sym}},N}^{\left( g_{N,r}^{- 1}(i) \right)}\left( y_{0}^{N - 1}\Pi_{g_{N,r}^{- 1}},\left\lbrack \mathbf{0}_{N/2},u_{\Lambda_{r}^{i - 1}} \right\rbrack \middle| u_{i} \right)
\end{align}
where \(\Lambda_{r}^{i - 1}\) denotes
\(\Lambda_{r} \cap \left\{ 0,\ldots,i - 1 \right\}\), \(\mathbf{0}_{N/2}\) represents the \(N/2\)-length all-zero sequence, and \(\Pi_{g_{N,r}^{- 1}}\) refers to the permutation matrix determined by \(g_{N,r}^{-1}\), i.e., \(\Pi_{g_{N,r}^{- 1}}(i,j)=!(j-g_{N,r}^{-1}(i))\).

The proof of Theorem 5 is in Appendix B.

Theorem 5 implies that by permuting \(y_{0}^{N - 1}\) with \(\Pi_{g_{N,r}^{- 1}}\),
the information bits \(u_{\Lambda_{r}}\), i.e., \(c_{0}^{K - 1}\) could
be decoded by a symmetric SCL decoder with information index set
\(g_{N,q}^{-1}(\mathcal{A})\).

In conclusion, due to the capacity discrepancy caused by frozen bits,
indices within \(\Lambda_{r}\) could no longer be selected according to
their symmetric capacities, instead, according to Theorem 4, by
selecting indices with highest sub-channel symmetric capacities within
\(\left\{ N\text{/}2,\ldots,N - 1 \right\}\) and mapping them into
\(\Lambda_{r}\) by \(g_{N,r}\), the indices with highest sub-channel
CIS-constrained capacities within \(\Lambda_{r}\) are determined. Moreover, in
decoding, according to Theorem 5, in order to calculate the CIS-constrained transition probability, rather than
the symmetric one, \(y_{0}^{N - 1}\) should be permuted and information
index set of the SCL decoder should be
\(g_{N,q}^{- 1}\left( \mathcal{A} \right)\).

By this means, we propose \textit{CIS-constrained construction and decoding} (\textit{CCD}) for comb-shaping polar codes, which still follows the system model presented in Fig.~\ref{fig_sys} but further specifies that:

1) In the transmitter, select the information index set \(\mathcal{A}\) by formula (\ref{eq_sel}).

2) In the receiver, permute
\(y_{0}^{N - 1}\) by \(\Pi_{g_{N,r}^{-1}}\):
\begin{align}
	{\widetilde{y}}_{0}^{N - 1} = y_{0}^{N - 1}\Pi_{g_{N,r}^{-1}}
\end{align}
and then send \({\widetilde{y}}_{0}^{N - 1}\) to an SCL decoder with information indices
\(g_{N,q}^{-1}(\mathcal{A})\).

The operation added by CCD is simply a permutation, consequently, the additional complexity of CCD is negligible.

} 

\section{Simulation Results}

\begin{figure*}[!t]
	\centering
	\subfloat[]{\includegraphics[width=3.25in]{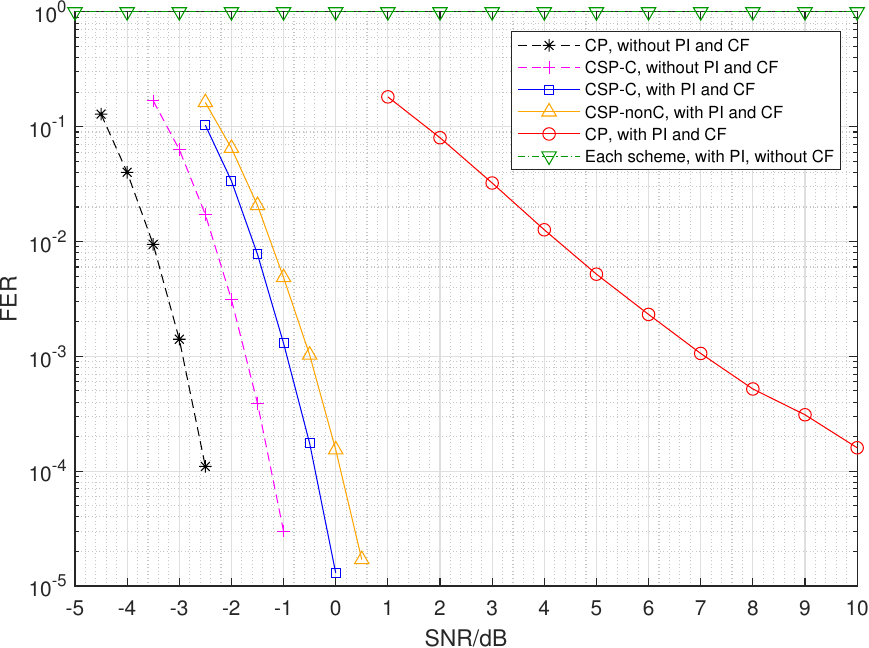}
		\label{fig_first_case}}
	\hfil
	\subfloat[]{\includegraphics[width=3.25in]{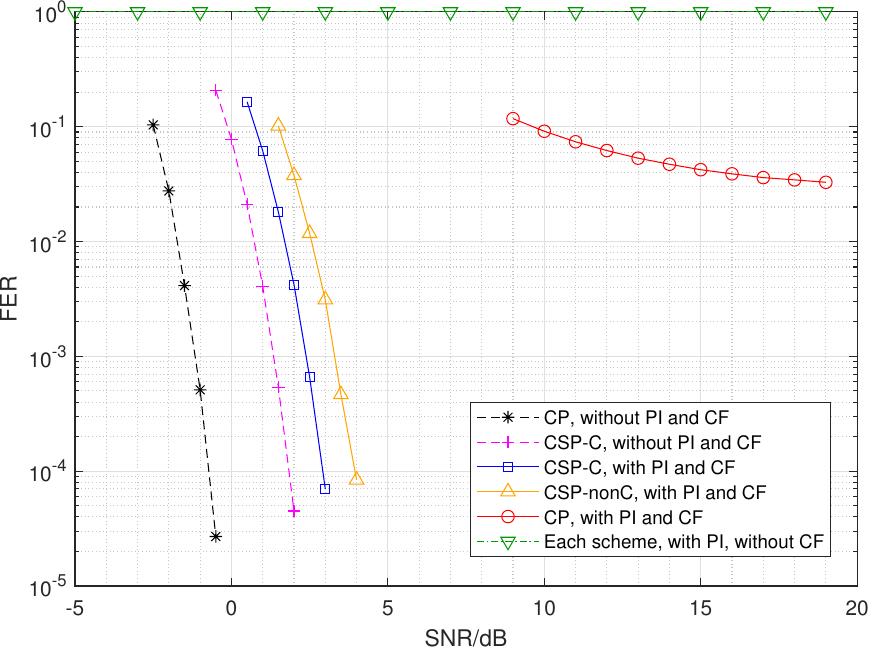}
		\label{fig_second_case}}
	\caption{\textcolor{black}{Frame error rates under code length 256. CP stands for the conventional polar code; CSP stands for the comb-shaping polar code, and CSP-C stands for CSP employing CCD, CSP-nonC stands for CSP not employing CCD. PI stands for periodic interference with fundamental frequency 50Hz, tone-bandwidth 20Hz and SIR -20dB. CF stands for a comb filter. (a) Code rate is 1/4; (b) Code rate is 3/8.}}
	\label{fig_fers256}
\end{figure*}

\begin{figure*}[!t]
	\centering
	\subfloat[]{\includegraphics[width=3.25in]{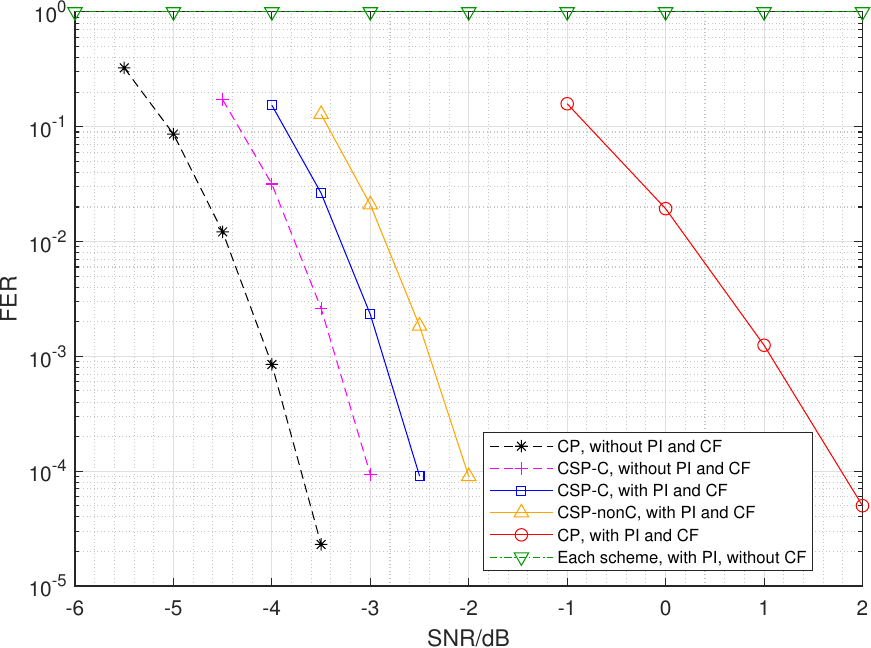}
		\label{fig_first_case}}
	\hfil
	\subfloat[]{\includegraphics[width=3.25in]{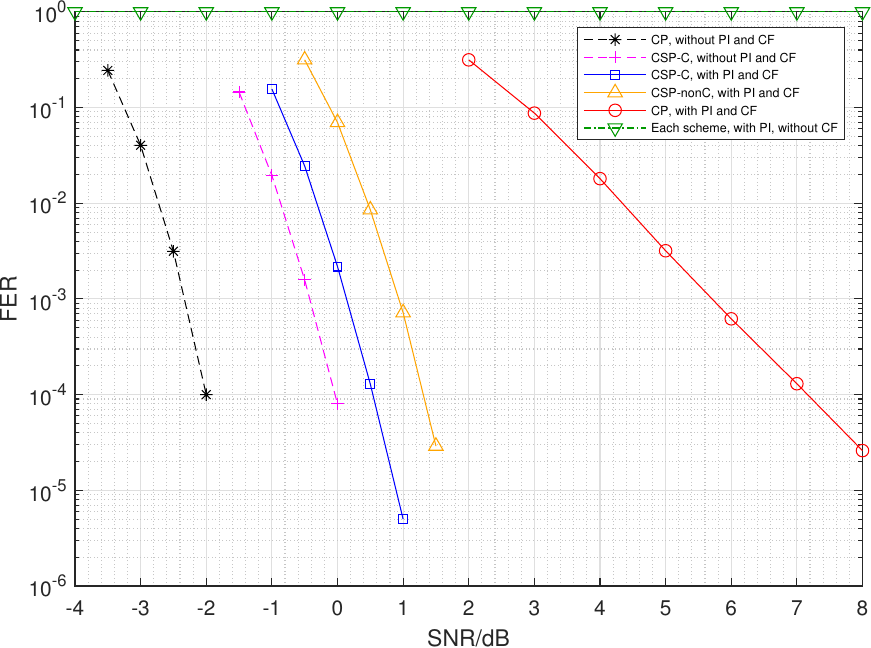}
		\label{fig_second_case}}
	\caption{\textcolor{black}{Frame error rates under code length 1024. CP stands for the conventional polar code; CSP stands for the comb-shaping polar code, and CSP-C stands for CSP employing CCD, CSP-nonC stands for CSP not employing CCD. PI stands for periodic interference with fundamental frequency 50Hz, tone-bandwidth 20Hz and SIR -20dB. CF stands for a comb filter. (a) Code rate is 1/4; (b) Code rate is 3/8.}}
	\label{fig_fers1024}
\end{figure*}

Having established the theoretical framework for both comb-shaping and performance enhancement, we now turn to numerical simulations to validate the effectiveness of our proposed schemes in practical scenarios.
The modulation is BPSK
with symbol rate \(R_{s} = 800\mathrm{Hz}\), sampling rate \(f_{s} = 6.4\mathrm{kHz}\),
and the shaping filter is an SRRC filter with roll-off factor 0.25,
period number 2 and sampling factor 8. The channel contains AWGN
noise and periodic interference. The signal-to-noise power ratio (SNR)
of the Gaussian noise is defined as the power ratio of the signal to the
noise within the signal band. The periodic interference has a
fundamental frequency 50Hz and tone-bandwidth 20Hz, whose expression is given in formula (\ref{eq_actItf}), \textcolor{black}{namely 40\% bandwidth is under interference}. The signal-to-interference power ration (SIR), similarly to
the SNR, is defined as the power ratio of the signal to the interference
within the signal band, and is set as -20dB, \textcolor{black}{which represents a rather strong interference}. \textcolor{black}{The comb filter has spectral notches centered at each interference frequency and a notch bandwidth 20Hz.} The number of decoding paths of the SCL decoder is 8. \textcolor{black}{The decoder is a conventional SCL decoder, so its complexity and memory usage are identical with SCL decoder \cite{ref:SCL}.}

Fig.~\ref{fig_fers256}-\ref{fig_fers1024} show the frame error rate (FER) of the comb-shaping and
conventional polar code under different conditions. Fig.~\ref{fig_fers256} shows the results under code length 256 and code rates 1/4, 3/8, the CIS chosen is
\(\Lambda_{3}\). \textcolor{black}{Fig.~\ref{fig_fers1024} shows the results under code length 1024 and code rates 1/4, 3/8, the CIS chosen is
\(\Lambda_{5}\).}

Evidently in Fig.~\ref{fig_fers256}-\ref{fig_fers1024}, under periodic interference, a remarkable performance advantage over the conventional polar code is achieved by the comb-shaping polar code. \textcolor{black}{Notably, under shorter code lengths and higher code rates, the error floor of conventional polar code becomes more evident, but this drawback is overcome by comb-shaping polar code. Furthermore, comparing Fig.~\ref{fig_fers256}(a) to Fig.~\ref{fig_fers256}(b), and Fig.~\ref{fig_fers1024}(a) to Fig.~\ref{fig_fers1024}(b), we find that under higher code rates, the advantage of the comb-shaping polar codes over the conventional ones is more significant; moreover, comparing Fig.~\ref{fig_fers256}(a) to Fig.~\ref{fig_fers1024}(a), and Fig.~\ref{fig_fers256}(b) to Fig.~\ref{fig_fers1024}(b), we find that under shorter code lengths, the advantage of the comb-shaping polar codes over the conventional ones is more significant. Moreover, with CCD employed in comb-shaping polar codes, the advantage is larger.}

\section{Conclusion}

In this paper, we proposed a novel spectral comb shaping scheme for polar codes to combat periodic interference. Moving beyond the traditional approach of passive filtering, which inevitably causes signal distortion, our method proactively shapes the signal spectrum to be separable from interference. By selecting information indices within a CIS proposed in this paper, periodic nulls and notch bands are introduced in the signal spectrum. Further with appropriate parameters, the signal spectral notches
could cover the all interference tones. That means
the periodic interference could be removed without much damage on the signal.
Besides, by employing CCD, the discrepancy in capacity and transition probability are solved, thus enhancing the error performance of comb-shaping polar code under AWGN. Numerical results confirm that our scheme significantly outperforms conventional methods in noisy environments with periodic interference. Furthermore, despite that the modulation considered in this paper is BPSK, according to the methodology of this research, our scheme could be readily extended to more kinds of linear modulations. Due to the length limit, this extension is left to future works.

\section{Appendix}

{\color{black}
\subsection{Proof of Theorem 4}

We first propose Lemma 3:

\emph{Lemma 3:} For any \(N=2^m, m\in\mathbb{N}^*\) and \( r \in \{0,\ldots,m-1\}\), there is \(\Pi_{g_{N,r}}G_N\Pi_{g_{N,r}^{- 1}} = G_N\).

\emph{Proof of Lemma 3:} According to the definition of \(\Pi_{g_{N,r}}\), there is
\(\left( \Pi_{g_{N,r}}G_N \right)(i,j) = \sum_{k = 0}^{N - 1}{\Pi_{g_{N,r}}(i,k)G_N(k,j)} = \\ G_N\left( g_{N,r}(i),j \right)\), and
\(\left( G_N\Pi_{g_{N,r}} \right)(i,j) = \sum_{k = 0}^{N - 1}{G_N(i,k)\Pi_{g_{N,r}}(k,j)} = G_N\left( i,g_{N,r}^{- 1}(j) \right)\).

Define function \(\eta_{N,r}:\left\{ 0,\ldots,m - 1 \right\} \mathord{\mapsto} \left\{ 0,\ldots,m - 1 \right\}\) by
\(\eta_{N,r}(d) :=
\left\{
\begin{aligned}
	& d  ,d \in \left\{ 0,\ldots,r-1 \right\} \\
	& d + 1  ,d \in \left\{ r,\ldots,m-2\right\} \\
	& r  ,d = m-1
\end{aligned} \right.\)
, then it is trivial that \(g_{N,r}(i) = \sum_{d = 0}^{m - 1}{i_{d}2^{\eta_{N,r}(d)}}\), namely \(g_{N,r}(i)_{\eta(d)} = i_{d}\), and that \(\eta_{N,r}\) is invertible, \(g_{N,r}^{-1}(i)_{\eta_{N,r}^{-1}(d)} = i_{d}\).
Further according to formula (\ref{eq_eleGN}), for any
\(i,j \in \left\{ 0,\ldots,N - 1 \right\}\ \)there is
\begin{align*}
	&G_N\left( g_{N,r}(i),j \right) \\
	&= ! |_{d = 0}^{m - 1}\left( ! g_{N,r}(i)_{d}\& j_{m-d-1} \right) = ! |_{d = 0}^{m - 1}\left( ! i_{\eta_{N,r}^{-1}(d)}\& j_{m-d-1} \right) \\
	&= ! |_{d = 0}^{m - 1}\left( ! i_{d}\& j_{\eta_{N,r}(m-d-1)} \right) = ! |_{d = 0}^{m - 1}\left( ! i_{d}\& g_{N,r}^{- 1}(j)_{m-d-1} \right) \\
	&= G_N\left( i,g_{N,r}^{- 1}(j) \right)
\end{align*}
i.e.,
\(\left( \Pi_{g_{N,r}}G_N \right)(i,j) \mathord{=} \left( G_N\Pi_{g_{N,r}} \right)(i,j),\forall i,j \mathord{\in} \left\{ 0,\ldots,N \mathord{-} 1 \right\}\),
briefly \(\Pi_{g_{N,r}}G_N = G_N\Pi_{g_{N,r}}\), i.e., \(\Pi_{g_{N,r}}G_N\Pi_{g_{N,r}^{- 1}} = G_N\). \(\hfill \blacksquare\)

With Lemma 3 it is ready to prove Lemma 4:

\emph{Lemma 4:} \(\forall m \in \mathbb{N}^*, r \in \{0,\ldots,m-1\}, i \in \Lambda_r, I_{r,N}^{(i)} = I_{m-1,N}^{(g_{N,r}^{-1}(i))}\), where \(N=2^m\).

\emph{Proof of Lemma 4:} For \(U_{\Lambda_{r}^{\complement}} \mathord{=} \lbrack 0,\ldots,0\rbrack,U_{\Lambda_{r}} \mathord{\sim} \mathcal{B}^{N/2}\left( \frac{1}{2} \right), \\ Z_{0}^{N - 1} \mathord{\sim} \mathcal{CN}^N (0,2\sigma^2), Y_{0}^{N - 1} = q\left( U_{0}^{N - 1}G_{N} \right) + Z_{0}^{N - 1}\), according to the definition of \(I_{r,N}^{(i)}\), we know \( I\left( Y_{0}^{N - 1}U_{0}^{i - 1};U_{i} \right) \mathord{=} I_{r,N}^{(i)} \). Define \({\widetilde{Y}}_{0}^{N - 1}\mathord{:=}Y_{0}^{N - 1}\Pi_{g_{N,r}^{- 1}}\), \({\widetilde{U}}_{0}^{N - 1}\mathord{:=}U_{0}^{N - 1}\Pi_{g_{N,r}^{- 1}}, {\widetilde{Z}}_{0}^{N - 1} \mathord{:=} Z_{0}^{N - 1}\Pi_{g_{N,r}^{-1}}\), then 
\begin{align*}
	&{\widetilde{Y}}_{0}^{N - 1} = \left\lbrack q\left( {\widetilde{U}}_{0}^{N - 1} \Pi_{g_{N,r}} G_N \right) + Z_{0}^{N - 1} \right\rbrack\Pi_{g_{N,r}^{-1}} \\
	&= q\left( {\widetilde{U}}_{0}^{N - 1} \Pi_{g_{N,r}} G_N \Pi_{g_{N,r}^{-1}} \right) + {\widetilde{Z}}_{0}^{N - 1}
\end{align*}

According to Lemma 3, \(\Pi_{g_{N,r}}G_N\Pi_{g_{N,r}^{- 1}} = G_N\), so there is
\({\widetilde{Y}}_{0}^{N - 1} = q\left( {\widetilde{U}}_{0}^{N-1} G_N \right) + {\widetilde{Z}}_{0}^{N - 1}\).

Note that \({\widetilde{U}}_{N/2}^{N-1} = \mathbf{0}_{N/2}, {\widetilde{U}}_{N/2}^{N-1} \sim \mathcal{B}^{N/2}(\frac{1}{2})\), and \({\widetilde{Z}}_{0}^{N - 1}\) is a permutation of AWGN \(Z_{0}^{N - 1}\), so \({\widetilde{Z}}_{0}^{N - 1} \sim \mathcal{CN}^N (0,2\sigma^2)\), therefore, according to the definition of \(I_{m-1,N}^{(i)}\), there is \(I({\widetilde{Y}}_{0}^{N - 1} \widetilde{U}_{0}^{i-1} ; \widetilde{U}_i)=I_{m-1,N}^{(i)},\forall i \in \{0,\ldots,N-1\}\).

From the definition of \(g_{N,r}^{-1}\), it could be readily shown that for each \(i\mathord{\in}\Lambda_r\), there is \({\widetilde{U}}_{0}^{g_{N,r}^{- 1}(i) - 1} \mathord{=} \left\lbrack \mathbf{0}_{N/2}, U_{\Lambda_{r}^{i - 1}} \right\rbrack\),
trivially
\(U_{0}^{i - 1} \mathord{\mapsto} {\widetilde{U}}_{0}^{g_{N,r}^{- 1}(i) - 1}\) is
invertible; and \({\widetilde{U}}_{g_{N,r}^{- 1}(i)} = U_{i}\), therefore
\(I\left( Y_{0}^{N - 1}U_{0}^{i - 1};U_{i} \right) \mathord{=} I\left( {\widetilde{Y}}_{0}^{N - 1}{\widetilde{U}}_{0}^{g_{N,r}^{- 1}(i) - 1};{\widetilde{U}}_{g_{N,r}^{- 1}(i)} \right)\). Recalling \(I({\widetilde{Y}}_{0}^{N - 1} \widetilde{U}_{0}^{i-1} ; \widetilde{U}_i)=I_{m-1,N}^{(i)},\forall i \in \{0,\ldots,N-1\}\), there is \(I\left( Y_{0}^{N - 1}U_{0}^{i - 1};U_{i} \right) = I_{m-1,N}^{(g_{N,r}^{-1}(i))}\), i.e., \(I_{r,N}^{(i)} = I_{m-1,N}^{(g_{N,r}^{-1}(i))}\).
\(\hfill \blacksquare\)

In addition, we prove Lemma 5:

\emph{Lemma 5:} For any mapping \(q:\mathbb{B \mapsto C}\), there is
\(\forall u,v \in \mathbb{B,}(1 - 2v)\left( q(u) - q_{a} \right) + q_{a} = \ q(u + v)\),
where \(q_{a} := \frac{1}{2}\left( q(0) + q(1) \right)\).

\emph{Proof of Lemma 5:} If \(v = 0\), then
\((1 - 2v)\left( q(u) - q_{a} \right) + q_{a} = q(u) = q(u + v)\); and
if \(v = 1\), then
\((1 - 2v)\left( q(u) - q_{a} \right) + q_{a} = 2q_{a} - q(u) = q(0) + q(1) - q(u) = q(!u) = q(u + v)\).
In summary,
\(\forall u,v \in \mathbb{B,\ }(1 - 2v)\left( q(u) - q_{a} \right) + q_{a} = q(u + v)\).
\(\hfill \blacksquare\)

With Lemma 5 it is ready to prove Lemma 6:

\emph{Lemma 6:} \(\forall m \in \mathbb{N}^*, i \in \{N/2,\ldots,N-1\}, I_{m-1,N}^{(i)} = I_{\mathrm{sym},N}^{(i)}\), where \(N=2^m\).

\emph{Proof of Lemma 6:} For \nolinebreak
\(U_{0}^{N - 1} \mathord{\sim} \mathcal{B}^{N}\left( \frac{1}{2} \right), Z_{0}^{N - 1} \mathord{\sim} \mathcal{CN}^{N}\left( 0,2\sigma^{2} \right),\\ Y_{0}^{N - 1} \mathord{=} q\left( U_{0}^{N - 1}G_{N} \right) \mathord{+} Z_{0}^{N - 1}\),
according to the definition of \(I_{\mathrm{sym},N}^{(i)}\), we know
\(I\left( Y_{0}^{N - 1}U_{0}^{i - 1};U_{i} \right) = I_{\mathrm{sym},N}^{(i)}\).
Define \nolinebreak
\({\widehat{U}}_{0}^{N - 1} \mathord{:=} \left\lbrack \mathbf{0}_{N/2},U_{N/2}^{N - 1} \right\rbrack, {\overline{U}}_{0}^{N - 1} \mathord{:=} U_{0}^{N - 1} \mathord{+} {\widehat{U}}_{0}^{N - 1} \mathord{=}\\ \left\lbrack U_{0}^{N/2 - 1},\mathbf{0}_{N/2} \right\rbrack\),
and
\({\widehat{Y}}_{0}^{N - 1} \mathord{:=} \left( 1 - 2{\overline{U}}_{0}^{N - 1}G_{N} \right)\left( Y_{0}^{N - 1} - q_{a} \right) \\ \mathord{+} q_{a}\), \({\widehat{Z}}_{0}^{N - 1} \mathord{:=} \left( 1 - 2{\overline{U}}_{0}^{N - 1} \right){\overline{Z}}_{0}^{N - 1}\), then \({\widehat{Y}}_{0}^{N - 1} = \left( 1 - 2{\overline{U}}_{0}^{N - 1}G_{N} \right)\left( q\left( U_{0}^{N - 1}G_{N} \right) - q_{a} \right) + q_{a}
+ {\widehat{Z}}_{0}^{N - 1}\),
according to Lemma 5 there is
\begin{align*}
	&{\widehat{Y}}_{0}^{N - 1}
	= q\left( U_{0}^{N - 1}G_{N} + {\overline{U}}_{0}^{N - 1}G_{N} \right) + {\widehat{Z}}_{0}^{N - 1} \\
	&= q\left( {\widehat{U}}_{0}^{N - 1}G_{N} \right) + {\widehat{Z}}_{0}^{N - 1}
\end{align*}
from where we know \({\widehat{Y}}_{0}^{N - 1}\) is independent of
\({\overline{U}}_{0}^{N - 1}\).

For each \(i \in \{ N/2,\ldots,N - 1\}\),
\(U_{0}^{i - 1} = \left\lbrack U_{0}^{N/2 - 1},U_{N/2}^{i - 1} \right\rbrack\),
note that \(U_{0}^{N/2 - 1} \mapsto {\overline{U}}_{0}^{N - 1}\)
and \(U_{N/2}^{i - 1} \mapsto {\widehat{U}}_{0}^{i - 1}\) are
invertible, then
\(U_{0}^{i - 1} \mapsto {\overline{U}}_{0}^{N - 1}{\widehat{U}}_{0}^{i - 1}\)
is invertible. Moreover, noting that
\(Y_{0}^{N - 1} = \left( 1 - 2{\overline{U}}_{0}^{N - 1}G_{N} \right)\left( {\widehat{Y}}_{0}^{N - 1} - q_{a} \right) + q_{a}\),
we know
\({\widehat{Y}}_{0}^{N - 1}{\overline{U}}_{0}^{N - 1} \mapsto Y_{0}^{N - 1}{\overline{U}}_{0}^{N - 1}\)
is invertible. Therefore, we have
\begin{align*}
	&I\left( Y_{0}^{N - 1}U_{0}^{i - 1};U_{i} \right) = I\left( Y_{0}^{N - 1}{\overline{U}}_{0}^{N - 1}{\widehat{U}}_{0}^{i - 1};{\widehat{U}}_{i} \right) \\
	&= I\left( {\widehat{Y}}_{0}^{N - 1}{\overline{U}}_{0}^{N - 1}{\widehat{U}}_{0}^{i - 1};{\widehat{U}}_{i} \right) = I\left( {\widehat{Y}}_{0}^{N - 1}{\widehat{U}}_{0}^{i - 1}{\overline{U}}_{0}^{N - 1};{\widehat{U}}_{i} \right)
\end{align*}

Here \({\overline{U}}_{0}^{N - 1}\) is independent of
\({\widehat{Y}}_{0}^{N - 1}{\widehat{U}}_{0}^{i - 1}\) and
\({\widehat{U}}_{i}\), so
\(I\left( {\overline{U}}_{0}^{N - 1};{\widehat{U}}_{i}|{\widehat{Y}}_{0}^{N - 1}{\widehat{U}}_{0}^{i - 1} \right) = 0\).
Therefore, by the chain rule, we have
\(I\left( {\widehat{Y}}_{0}^{N - 1}{\widehat{U}}_{0}^{i - 1}{\overline{U}}_{0}^{N - 1};{\widehat{U}}_{i} \right) = I\left( {\widehat{Y}}_{0}^{N - 1}{\widehat{U}}_{0}^{i - 1};{\widehat{U}}_{i} \right) + I\left( {\overline{U}}_{0}^{N - 1};{\widehat{U}}_{i}|{\widehat{Y}}_{0}^{N - 1}{\widehat{U}}_{0}^{i - 1} \right) = I\left( {\widehat{Y}}_{0}^{N - 1}{\widehat{U}}_{0}^{i - 1};{\widehat{U}}_{i} \right)\).
Briefly,
\(I\left( Y_{0}^{N - 1}U_{0}^{i - 1};U_{i} \right) = I\left( {\widehat{Y}}_{0}^{N - 1}{\widehat{U}}_{0}^{i - 1};{\widehat{U}}_{i} \right)\).

Noting that
\({\widehat{U}}_{0}^{N/2 - 1} \mathord{=} \lbrack 0,\ldots,0\rbrack,{\widehat{U}}_{N/2}^{N - 1} \mathord{\sim} \mathcal{B}^{N/2}\left( \frac{1}{2} \right), {\widehat{Z}}_{0}^{N - 1} \sim \mathcal{C}\mathcal{N}^{N}\left( 0,2\sigma^{2} \right),\ {\widehat{Y}}_{0}^{N - 1} = q\left( {\widehat{U}}_{0}^{N - 1}G_{N} \right) + {\widehat{Z}}_{0}^{N - 1}\),
according to the definition of \(I_{m - 1,N}^{(i)}\), there is
\(I\left( {\widehat{Y}}_{0}^{N - 1}{\widehat{U}}_{0}^{i - 1};{\widehat{U}}_{i} \right) = I_{m - 1,N}^{(i)}\).
Therefore,
\(I\left( Y_{0}^{N - 1}U_{0}^{i - 1};U_{i} \right) = I_{m - 1,N}^{(i)}\),
i.e.,
\(I_{\mathrm{sym},N}^{(i)} = I_{m - 1,N}^{(i)},\forall i \in \{ N/2,\ldots,N - 1\}\).
\(\hfill \blacksquare\)

With Lemma 4 and 6 it is ready to prove Theorem 4:

\emph{Proof of Theorem 4:} \(\forall m \in \mathbb{N}^*, r \in \{0,\ldots,m-1\}, i \in \Lambda_r\), according to Lemma 4,  \(I_{r,N}^{(i)} = I_{m-1,N}^{(g_{N,r}^{-1}(i))}\), where \(N=2^m\). Note that \(g_{N,r}^{-1}(i) \in \{N/2,\ldots,N-1\}\), then according to Lemma 6, there is \(I_{m-1,N}^{(g_{N,r}^{-1}(i))} = I_{\mathrm{sym},N}^{(g_{N,r}^{-1}(i))}\). In conclusion, there is \(I_{r,N}^{(i)} = I_{\mathrm{sym},N}^{(g_{N,r}^{-1}(i))}\). \(\hfill \blacksquare\)

\subsection{Proof of Theorem 5}

For \(U_{\Lambda_{r}^{\complement}} \mathord{=} \lbrack 0,\ldots,0\rbrack,U_{\Lambda_{r}} \mathord{\sim} \mathcal{B}^{N/2}\left( \frac{1}{2} \right), Z_{0}^{N - 1} \mathord{\sim} \mathcal{CN}^N (0,2\sigma^2), \\ Y_{0}^{N - 1} = q\left( U_{0}^{N - 1}G_{N} \right) + Z_{0}^{N - 1}\), according to the definition of \(W_{r,N}^{(i)}\), we know \( p_{Y_{0}^{N - 1}U_{\Lambda_{r}^{i - 1}}|U_{i}} \left( y_{0}^{N - 1},u_{\Lambda_{r}^{i - 1}}|u_{i} \right) = W_{r,N}^{(i)} \left( y_{0}^{N - 1}u_{\Lambda_{r}^{i - 1}}|u_{i} \right) \). Define \({\widetilde{Y}}_{0}^{N - 1}:=Y_{0}^{N - 1}\Pi_{g_{N,r}^{- 1}}, {\widetilde{U}}_{0}^{N - 1}:=U_{0}^{N - 1}\Pi_{g_{N,r}^{- 1}}, {\widetilde{Z}}_{0}^{N - 1} := Z_{0}^{N - 1}\Pi_{g_{N,r}^{-1}}\), then 
\begin{align*}
	&{\widetilde{Y}}_{0}^{N - 1} = \left\lbrack q\left( {\widetilde{U}}_{0}^{N - 1} \Pi_{g_{N,r}} G_N \right) + Z_{0}^{N - 1} \right\rbrack\Pi_{g_{N,r}^{-1}} \\
	&= q\left( {\widetilde{U}}_{0}^{N - 1} \Pi_{g_{N,r}} G_N \Pi_{g_{N,r}^{-1}} \right) + {\widetilde{Z}}_{0}^{N - 1}
\end{align*}

According to Lemma 3, \(\Pi_{g_{N,r}}G_N\Pi_{g_{N,r}^{- 1}} = G_N\), so there is
\({\widetilde{Y}}_{0}^{N - 1} = q\left( {\widetilde{U}}_{0}^{N-1} G_N \right) + Z_{0}^{N - 1}\Pi_{g_{N,r}^{-1}}\).
Define \({\widetilde{Z}}_{0}^{N - 1} = Z_{0}^{N - 1}\Pi_{g_{N,r}^{-1}}\), then there is
\({\widetilde{Y}}_{0}^{N - 1} = q\left( {\widetilde{U}}_{0}^{N-1} G_N \right) + {\widetilde{Z}}_{0}^{N - 1}\).

Note that \({\widetilde{U}}_{N/2}^{N-1} = \mathbf{0}_{N/2}, {\widetilde{U}}_{N/2}^{N-1} \sim \mathcal{B}^{N/2}(\frac{1}{2})\), and \({\widetilde{Z}}_{0}^{N - 1}\) is a permutation of AWGN \(Z_{0}^{N - 1}\), so \({\widetilde{Z}}_{0}^{N - 1} \sim \mathcal{CN}^N (0,2\sigma^2)\), therefore, according to the definition of \(W_{m-1,N}^{(i)}\), there is \(p_{{\widetilde{Y}}_{0}^{N - 1} \widetilde{U}_{N/2}^{i-1}|\widetilde{U}_i} ({\widetilde{y}}_{0}^{N - 1} \widetilde{u}_{N/2}^{i-1}|\widetilde{u}_i)=W_{m-1,N}^{(i)} ({\widetilde{y}}_{0}^{N - 1} \widetilde{u}_{N/2}^{i-1}|\widetilde{u}_i),\forall i \in \{N/2,\ldots,N-1\}\).

From the definition of \(g_{N,r}^{-1}\), it could be readily shown that for each \(i\mathord{\in}\Lambda_r\), there is \({\widetilde{U}}_{N/2}^{g_{N,r}^{- 1}(i) - 1} \mathord{=} U_{\Lambda_{r}^{i - 1}}\) and \({\widetilde{U}}_{g_{N,r}^{- 1}(i)} = U_{i}\), therefore
\(p_{Y_{0}^{N - 1}U_{\Lambda_{r}^{i - 1}}|U_{i}} \left( y_{0}^{N - 1},u_{\Lambda_{r}^{i - 1}}|u_{i} \right) = p_{{\widetilde{Y}}_{0}^{N - 1}{\widetilde{U}}_{N/2}^{g_{N,r}^{- 1}(i) - 1}|{\widetilde{U}}_{g_{N,r}^{- 1}(i)}} \left( y_{0}^{N - 1} \Pi_{g_{N,r}^{- 1}},u_{\Lambda_{r}^{i - 1}} \middle| u_i \right)\). Recalling \(p_{{\widetilde{Y}}_{0}^{N - 1} \widetilde{U}_{N/2}^{i-1}|\widetilde{U}_i} ({\widetilde{y}}_{0}^{N - 1} \widetilde{u}_{N/2}^{i-1}|\widetilde{u}_i) \mathord{=} W_{m-1,N}^{(i)} ({\widetilde{y}}_{0}^{N - 1} \widetilde{u}_{N/2}^{i-1}|\widetilde{u}_i),\forall i \in \{N/2,\ldots,N\mathord{-}1\}\), there is \nolinebreak \(p_{Y_{0}^{N - 1}U_{\Lambda_{r}^{i - 1}}|U_{i}} \left( y_{0}^{N - 1},u_{\Lambda_{r}^{i - 1}}|u_{i} \right) \\ \mathord{=} W_{m-1,N}^{(g_{N,r}^{-1}(i))} \left( y_{0}^{N - 1} \Pi_{g_{N,r}^{- 1}},u_{\Lambda_{r}^{i - 1}} \middle| u_i \right)\), in other words, \(W_{r,N}^{(i)} \left( y_{0}^{N - 1},u_{\Lambda_{r}^{i - 1}}|u_{i} \right) \mathord{=} W_{m-1,N}^{(g_{N,r}^{-1}(i))} \left( y_{0}^{N - 1} \Pi_{g_{N,r}^{- 1}},u_{\Lambda_{r}^{i - 1}} \middle| u_i \right)\).

Trivially, \(W_{m-1,N}^{(g_{N,r}^{-1}(i))} \left( y_{0}^{N - 1} \Pi_{g_{N,r}^{- 1}},u_{\Lambda_{r}^{i - 1}} \middle| u_i \right) = 2^{N/2} W_{\mathrm{sym},N}^{(g_{N,r}^{-1}(i))} \left( y_{0}^{N - 1} \Pi_{g_{N,r}^{- 1}},\left\lbrack \mathbf{0}_{N/2}, u_{\Lambda_{r}^{i - 1}} \right\rbrack \middle| u_i \right)\), therefore, there is \(W_{r,N}^{(i)} \left( y_{0}^{N - 1},u_{\Lambda_{r}^{i - 1}}|u_{i} \right) = 2^{N/2}W_{\mathrm{sym},N}^{(g_{N,r}^{-1}(i))} \left( y_{0}^{N - 1} \Pi_{g_{N,r}^{- 1}},\left\lbrack \mathbf{0}_{N/2}, u_{\Lambda_{r}^{i - 1}} \right\rbrack \middle| u_i \right)\).
\(\hfill \blacksquare\)
} 

\end{document}